\documentclass[%
twocolumn,
superscriptaddress,
longbibliography,
nofootinbib,
amsmath,amssymb,
aps,
notitlepage
]{revtex4-1}

\usepackage{graphicx}
\usepackage{dcolumn}
\usepackage{bm}
\usepackage{hyperref}
\usepackage{color}
\usepackage{xcolor}
\usepackage{amsmath}
\usepackage{amssymb}
\usepackage{acronym}
\usepackage{xspace}
\usepackage{subfigure}
\usepackage{longtable} 
\usepackage{multirow}
\usepackage{mathtools}
\usepackage{soul}
\usepackage[normalem]{ulem}

\usepackage{orcidlink}

\newcommand{\IUCAA}{Inter-University Centre for Astronomy and Astrophysics, Post Bag 4, Ganeshkhind, Pune 411 007, India}
\newcommand{\ICRR}{Institute for Cosmic Ray Research, The University of Tokyo, Kashiwanoha 5-1-5, Kashiwa, Chiba 277-8582, Japan}
\newcommand{\WSU}{Department of Physics \& Astronomy, Washington State University, 1245 Webster, Pullman, WA 99164-2814, USA}
\newcommand{\OKC}{The Oskar Klein Centre, Department of Astronomy, Stockholm University, AlbaNova, SE-10691 Stockholm, Sweden}
\newcommand{\UHH}{Universit\"{a}t Hamburg, D-22761 Hamburg, Germany}

\newcommand{\kmsMpc}{\ensuremath{\mbox{km s}^{-1} \,\mbox{Mpc}^{-1}}\xspace}

\begin{document}

\title{Joint Inference of Population, Cosmology, and Neutron Star Equation of State from Gravitational Waves of Dark Binary Neutron Stars}

\author{Tathagata Ghosh~\orcidlink{0000-0001-9848-9905}}

\affiliation{\IUCAA}
\affiliation{\ICRR}

\author{Bhaskar Biswas~\orcidlink{0000-0003-2131-1476}}
\affiliation{\UHH} 
\affiliation{\OKC}

\author{Sukanta Bose~\orcidlink{0000-0002-4151-1347}}
\affiliation{\WSU}
\affiliation{\IUCAA}

\author{Shasvath J. Kapadia~\orcidlink{0000-0001-5318-1253}}
\affiliation{\IUCAA}

\begin{abstract}

Gravitational waves (GWs) from binary neutron stars (BNSs) are expected to be accompanied by electromagnetic (EM) emissions, which help identify the host galaxy. Since GWs directly measure their luminosity distances, joint GW--EM observations from BNSs help with the study of cosmology, particularly the Hubble constant, unaffected by cosmic distance ladder systematics. However, detecting the EM emissions is not always possible. Additionally, the tidal deformability of neutron stars (NSs), combined with the knowledge of the NS EoS, can break the degeneracy between mass parameters and redshift, allowing for the inference of the Hubble constant. While several studies have aimed to infer the Hubble constant using dark BNSs (without EM counterparts), none have consistently combined the uncertainties of population, cosmology, and NS EoS within a Bayesian framework. In this study, we propose a novel Bayesian analysis to jointly constrain the NS EoS, population, and cosmological parameters using a population of dark BNSs detected through GW observations. We demonstrate the statistical robustness of our method using $50$ simulated BNS events following Gaussian and double Gaussian mass distributions, detected by Advanced LIGO and Advanced Virgo detectors operating at O5 sensitivity. We show that such measurements can constrain the Hubble constant with a precision of $\lesssim 35\%$ ($90\%$ credible interval). This level of precision is unattainable without incorporating NS EoS, especially when observing BNS mergers without EM counterpart information. We also report the Hubble constant measurements obtained from a more realistic set of $5$ simulated BNS events.

\end{abstract}

\date{\today}

\maketitle

\section{Introduction}

In the present era of precision cosmology, one of the primary pursuits remains the precise measurement of the Hubble constant, $H_{0}$ -- a fundamental observable that quantifies the current expansion rate of the Universe. However, presently there exists a discrepancy between the measurements of the Hubble constant from different observations. 
The SH0ES Collaboration measured the Hubble constant to be $H_{0}=73.04\pm 1.04~\kmsMpc$ using $42$ Type Ia supernovae (SNe Ia) calibrated with Cepheid variable stars~\citep{Riess:2021jrx}.
More recently, an independent analysis of $16$ SNe Ia calibrated with JWST Cepheids yielded $H_{0}=72.6\pm 2~\kmsMpc$, 
which is consistent with the value expected from HST Cepheids for the same SNe Ia sample, viz. $H_{0}=72.8\pm 2~\kmsMpc$~\citep{Riess:2024vfa}.
These late-Universe measurements are concordant but remain in $\sim 5\sigma$ tension with the indirect inference of $H_{0}=67.36~\pm 0.54~\kmsMpc$~\citep{Planck:2018vyg} from the Planck measurements of the cosmic microwave background.
This discordance in the measurement of the expansion rate of the Universe may either imply new physics beyond the standard model or be a systematic effect owing to unaccounted uncertainties in one or more measurements. However, there is no conclusive consensus regarding the value of the Hubble constant today. So, it is crucial to pursue observations independent of the aforementioned ones that can help resolve this tension in the values of the Hubble constant.

The first detection of gravitational waves (GWs) from the binary neutron star (BNS)
GW170817~\citep{LIGOScientific:2017vwq}
by LIGO~\citep{LIGOScientific:2014pky} and Virgo~\citep{VIRGO:2014yos} detectors -- along with 
the observations of its electromagnetic (EM) counterparts across the spectrum~\citep{LIGOScientific:2017ync} -- has opened the window to multimessenger astronomy involving GWs. This inter alia allows for probing the cosmic expansion of the Universe.
The independent measurements of the luminosity distance from GW data and the redshift from EM data, both from GW170817, 
enabled us to measure the Hubble constant to be $H_{0}=70^{+12.0}_{-8.0}~\kmsMpc$~\citep{LIGOScientific:2017adf}.
However, GW observations today are dominated by binary black hole (BBH) mergers~\citep{KAGRA:2021vkt}, which are not accompanied by EM counterparts.
As a result, the so-called bright siren approach, which relies on 
inferring the redshift of the binaries from the EM spectra of their galactic hosts, cannot be applied to those mergers.
So, alternative methods have been proposed to infer $H_{0}$ from GW observations without EM counterparts. These GW sources are referred to as dark sirens. These methods are primarily categorized into two approaches. One such method is the galaxy catalog method~\citep{Schutz:1986gp, DelPozzo:2011vcw, Gray:2019ksv, DES:2019ccw, Finke:2021aom, Gray:2021sew}, which statistically associates the dark sirens with the redshifts of galaxies within the sky localizations of the corresponding GW events, as potential host galaxies.
The other method, known as the spectral siren method~\citep{You:2020wju, Mastrogiovanni:2021wsd}, statistically derives the redshifts of GW events by comparing the source-frame mass spectrum with the observed masses of BBHs, which are redshifted, as explained later in this section. Both methods utilize complementary approaches to statistically infer the redshifts of GW events and, consequently, the Hubble constant.

In a recent study~\citep{LIGOScientific:2021aug}, the LIGO-Virgo-KAGRA (LVK) collaboration applied both the spectral siren and galaxy catalog methods to infer the Hubble constant to be $H_{0}=68^{+8}_{-6}$~\kmsMpc and $H_{0}=68^{+12}_{-8}$~\kmsMpc, respectively, by using $42$ BBHs from GWTC-3~\citep{KAGRA:2021vkt} as well as the BNS event GW170817. Each method has distinct strengths and weaknesses. The spectral siren method allows for the variation of GW population parameters but does not benefit from the additional constraining power of galaxy redshifts. In contrast, the galaxy catalog method fixes the GW population parameters during the computation of the selection function. This can introduce potential bias in estimating the Hubble constant if the assumed population model is incorrect.
So, there are ongoing efforts~\citep{Mastrogiovanni:2023emh, Gray:2023wgj} to unify these two methods: the galaxy catalog information is utilized to infer $H_{0}$ while marginalizing over the population model uncertainties. This approach effectively integrates the advantages of both methods. Furthermore, there is another approach --  termed the cross-correlation method -- that explores the expected clustering between GW sources and galaxies to infer the redshift information and hence the Hubble constant~\citep{Oguri:2016dgk,
Nair:2018ign, Bera:2020jhx, Mukherjee:2019wcg, Mukherjee:2020hyn, Diaz:2021pem, Mukherjee:2022afz, Ghosh:2023ksl}.

Similar to the spectral method applied to the BBH events, prior knowledge of the mass distribution of BNS also allows us to estimate the Hubble constant due to the narrowness of the NS mass spectrum~\citep{Taylor:2011fs, Taylor:2012db}.
The inspiral GW signal of a stellar-mass BBH is the result of point-particle orbital dynamics, without matter.
GW observables like the component masses and the signal frequency 
are covariant with the redshift factor $(1+z)$. In other words, it is not possible to extract their redshift from the measurement of their redshifted mass $m^{z} \equiv m(1+z)$, where $m$ is the source-frame mass of a component.
Unlike the black holes in binary mergers, neutron stars (NSs) are tidally deformed due to the presence of matter. This tidal deformation, which affects the GW phase evolution, depends on the source-frame masses. So, the measurement of tidal deformabilities~\footnote{As explored herein, the dimensionless tidal deformability is just a number and does not scale with $(1+z)$.} from the GW signal helps to break the mass-redshift degeneracy and was first proposed by Messenger \textit{et al.}~\cite{Messenger:2011gi}. Thus, the measurement of tidal deformabilities can constrain the distance-redshift relation and hence cosmological parameters, even in the absence of EM counterpart detections. 

Although only one or two BNS mergers have been confirmed so far,~\footnote{These are GW170817~\citep{LIGOScientific:2017vwq} and, with lesser confidence, GW190425~\citep{LIGOScientific:2020aai}.} nevertheless the existing LIGO-Virgo network, along with the addition of KAGRA and LIGO-India in the future, has the potential to detect a few to several hundred BNS mergers per year, up to redshift $z \sim 0.5$~\citep{KAGRA:2013rdx, Petrov:2021bqm, Pandey:2024mlo}. The next-generation GW observatories, such as the Einstein Telescope (ET) and Cosmic Explorer (CE), are expected to significantly increase detection rates -- potentially identifying up to $10^{5}$ BNS mergers per year -- as they extend the observable horizon to $z \sim 10$ with CE~\citep{Evans:2021gyd, Evans:2023euw} and $z \sim 4$ with ET~\citep{Branchesi:2023mws}. Motivated by this prospect for BNS detections, several efforts~\citep{Chatterjee:2021xrm, Ghosh:2022muc, Shiralilou:2022urk, Dhani:2022ulg} have been spawned by the idea of using BNS tidal deformabilities in GWs to measure the Hubble constant.
In Refs.~\cite{Chatterjee:2021xrm} and~\cite{Dhani:2022ulg}, the authors assume that the equation of state (EoS) is known exactly for the future-generation detector era.
On the other hand, Refs.~\cite{Ghosh:2022muc} and~\cite{Shiralilou:2022urk} allow for uncertainties in the NS EoS while inferring the Hubble constant in the future-generation detector era. Specifically, in Ref.~\cite{Ghosh:2022muc}, the BNS mass distribution is assumed to be uniform. Consequently, while the maximum mass is inferred from the constrained NS EoS, the minimum mass remains fixed. The choice of the mass model limits us from inferring other mass model parameters that are not explicitly dependent on the NS EoS. 
However, none of these works~\citep{Chatterjee:2021xrm, Ghosh:2022muc, Shiralilou:2022urk, Dhani:2022ulg} deduces the population model (including both mass and redshift distributions) of the underlying distribution of GW events. Notably, inferring the mass distribution in conjunction with the NS EoS is essential to mitigating bias in the mass distribution parameters~\citep{Wysocki:2020myz, Golomb:2021tll} and further constraining the NS EoS.

This paper demonstrates the simultaneous inference of mass distribution, redshift distribution, NS EoS, and the Hubble constant
from a population of BNSs, by unifying the corresponding model hyperparameters within a Bayesian framework.
This method becomes promising for future observations for several reasons. Firstly, it does not rely on EM counterparts (e.g., for obtaining the source redshift), which is necessary for measuring the Hubble constant.
Indeed, those counterparts may not be detectable for distant BNSs. Kilonovae, which are the outcome of BNS mergers, are not luminous enough to be detectable beyond a redshift of $z \sim 1$~\citep{Metzger:2011bv, Tanvir:2017pws}. EM counterparts in the form of short-hard gamma-ray bursts are detectable up to redshift $z \sim 2$~\citep{Paterson:2020rya}.
Large GW sky-localization regions do not make it easier to find EM counterparts either~\citep{ref:rana17,ref:rana19}.
Except for GW170817, none of the binary mergers involving NS(s) have been conclusively observed in EM~\citep{LIGOScientific:2020aai,LIGOScientific:2021qlt}.
In showing how to infer the population model,  cosmological parameters, and NS EoS solely from GW data -- even while employing uniform priors for EoS model parameters -- this methodology aims to go beyond earlier important
attempts~\citep{Chatterjee:2021xrm, Ghosh:2022muc, Shiralilou:2022urk, Dhani:2022ulg}.
However, note that in this work the signal simulations employ the same EoS and population models as those used by our Bayesian inference studies to assess the statistical errors in measuring the model hyperparameters. We leave the study of systematic errors to future work.

The paper is organized as follows. Sec.~\ref{sec:method} elaborates on the methodology used for inferring the population model, cosmological parameters, and EoS model. In Sec.~\ref{sec:models}, different parameterized models corresponding to NS EoS, mass, and redshift distribution are introduced. In Sec.~\ref{sec:simulation}, we discuss the simulation performed in this work to show the efficacy of the proposed methodology. The results are presented in Sec.~\ref{sec:results}. We conclude this paper with a discussion of the implications of our results for future GW detectors and nuclear experiments.

\section{Methodology} \label{sec:method}

In this section, we describe the Bayesian inference method employed for the joint estimation of EoS of NSs, along with their underlying mass distribution 
and redshift distribution using GW data from a population of BNS. The set of hyperparameters $\bm{\Theta} =  \{\bm{\Theta}_{\mathcal{E}}, \bm{\Theta}_{m}, \bm{\Theta}_{z}, \bm{\Theta}_{c}\}$ involved in this study corresponds to different parameterized models  
of NS EoS, BNS mass distribution, redshift evolution of the BNS merger rate, and cosmology (e.g., the value of $H_0$), respectively.
Specifically, we compare the estimates of population parameters and the Hubble constant
-- with and without NS EoS measurements -- to highlight the impact of EoS information in constraining those parameters. 
In the scenario where the NS EoS is not utilized, the set of hyperparameters is limited to $\bm{\Theta}=\{\bm{\Theta}_{m}, \bm{\Theta}_{z}, \bm{\Theta}_{c}\}$. 
In this case, measurement of the mass distribution proves useful in estimating $H_0$, and is essentially the spectral siren approach.
However, our Bayesian framework -- a generic approach -- remains the same whether or not the NS EoS measurements are utilized for inference.
We refer to these hyperparameters as model parameters for the rest of the paper.

In order to constrain the model parameters, one needs to first perform Bayesian estimation of the BNS parameters of all such detected events individually.~\footnote{A brief review of the Bayesian formalism used in this paper is given in Appendix~\ref{appendix_bayesian_formalism}.}
The posteriors of those parameters, once obtained, 
are subsequently utilized to infer joint posterior distributions of the model hyperparameters for $N$ events as follows:
\begin{equation}\label{eq:bayesian1}
    p(\bm{\Theta} \mid \{d\}) \propto p(\bm{\Theta}) \prod_{i=1}^{N} \frac{1}{\beta(\bm{\Theta})} \int \mathcal{L}(d_{i} \mid \bm{\theta}_{i}) p(\bm{\theta}_{i} \mid \bm{\Theta}) d\bm{\theta}_{i} \,,
\end{equation}
where $i$ is the event index, $d_i$ are the observational data for each event, $\{d\}$ is the set of data $d_i$ for all $N$ events, and 
$\bm{\theta} \equiv \{m_{1}, m_{2}, z, \Lambda_{1}, \Lambda_{2}\}$ is a set of BNS parameters.
To wit, these are the source-frame BNS masses $m_{1,2}$, their common redshift $z$, and their corresponding tidal deformabilities $\Lambda_{1,2}$.~\footnote{Note that  $z$ and $\bm{\Theta}_z$ are not the same. While the former is the redshift of an individual BNS, the latter is the redshift evolution hyperparameter of the BNS population.}
Here, $p(\bm{\Theta})$ denotes the prior distributions over all the hyperparameters and $p(\bm{\theta}_{i} \mid \bm{\Theta})$ represents the population model.
$\beta(\bm{\Theta})$ is a selection function that we describe later in this section.
Moreover, the likelihood $\mathcal{L}(d_{i} \mid \bm{\theta_{i}})$ of the individual events is constructed from the data by marginalizing over all other BNS parameters.~\footnote{The marginalized BNS parameters are, e.g., the orbit's inclination to the line of sight, right ascension and declination of the source, etc.}

It is important to note that the parameters of individual BNS events measured directly from GW data are detector-frame values, and are denoted by $\bm{\theta}_{d} = \{\mathcal{M}_{c}^{z}, q, d_{L}, \Lambda_{1}, \Lambda_{2}\}$, where $\mathcal{M}_{c}^{z}=\mathcal{M}_{c}(1+z)$ is the detector-frame chirp mass and $d_{L}(z, H_{0})$ is the luminosity distance. The mass ratio  $q=m_{2}/m_{1}$ and $\Lambda_{1,2}$, being pure numbers, are frame independent.
As is apparent, the source-frame values of the component masses and the chirp mass can be deduced from the detector-frame measurements if the source redshift is known or can be found, e.g., from the EM spectrum of the host galaxy.
Alternatively, if the NS EoS and mass distribution are known, the $m$–$\Lambda$ relation can be exploited to deduce the source-frame masses from the measured tidal deformabilities. Comparing the inferred source-frame masses with the respective detector-frame masses provides an estimate of the BNS redshift. Combined with the luminosity distance, this redshift can be used to constrain $H_0$.

The model parameters $\bm{\Theta}$ are most immediately associated with the source-frame parameters $\bm{\theta}$. Therefore, the posterior samples and the corresponding priors used to estimate the BNS parameters need to be converted from the detector frame, where the signal is measured, to the source frame.
Next, the source-frame posterior distributions should be divided by the corresponding source-frame priors $p_{\rm PE}(\bm{\theta}_{i})$ to construct the semimarginalized likelihood in the source frame, as follows:
\begin{equation} \label{likelihood}
    \mathcal{L}(d_{i} \mid \bm{\theta_{i}}) \propto \frac{p(\bm{\theta}_{i} \mid d_{i})}{p_{\rm PE}(\bm{\theta}_{i})} \,.
\end{equation}
Here, the single-event source-frame 
posterior $p(\bm{\theta}_{i} \mid d_{i})$ is constructed by applying the Gaussian mixture model~\citep{Talbot:2020oeu}~\footnote{This module is implemented in~\textsc{SCIKIT-LEARN}~\citep{2011JMLR...12.2825P}} to the BNS parameters $\bm{\theta}_{i}$ of the corresponding event.

It is important to note here that the source-frame priors ${p_{\rm PE}(\bm{\theta})}$  are not independent of the detector-frame priors $p_{\rm PE}(\bm{\theta}_{d})$, which would have been employed in estimating the signal parameters of individual BNS events. Indeed, the Jacobian of the ``coordinate" transformation relating $\bm{\theta}_{d}$ and $\bm{\theta}$ can be used to obtain the former prior from the latter as follows:
\begin{equation}
    p_{\rm PE}(\bm{\theta}) = \left\lvert J\left( \frac{\bm{\theta}_{d}}{\bm{\theta}} \right) \right\lvert \times p_{\rm PE}(\bm{\theta}_{d}) \,,
\end{equation}
where $\lvert J \rvert$ denotes the determinant of the Jacobian matrix $J$.
In this work, we have imposed uniform priors over detector-frame chirp mass ($\mathcal{M}_{c}^{z}$) and mass ratio ($q$) but a cosmology-independent $d_{L}^{2}$ prior over the luminosity distance. However, our analysis is based on the source-frame parameters. 
Note that the dimensionless tidal deformability parameter $\Lambda$ for any NS is the same in both frames; so we do not include it in the Jacobian matrix.
For the given prior probability defined on $(\mathcal{M}_{c}^{z}, q, d_{L})$, the corresponding probability in $(m_{1}, m_{2}, z)$ is~\citep{Callister:2021gxf}
\begin{eqnarray} \label{jacobian}
    p_{\rm PE}(m_{1}, m_{2}, z) && = p_{\rm PE}(\mathcal{M}_{c}^{z}, q, d_{L}) \left\lvert J \left( \frac{\mathcal{M}_{c}^{z}, q, d_{L}}{m_{1}, m_{2}, z}\right) \right\rvert \nonumber \\
    && \propto  d_{L}^{2} (1+z)  \frac{\partial d_{L}}{\partial z} \frac{(m_{1} m_{2})^{3/5}}{m_{1}^{2} (m_{1}+m_{2})^{1/5}} \,.
\end{eqnarray}
In Eq.~\eqref{jacobian}, the first factor in the last expression comes from the $d_{L}^{2}$ prior, and the rest of the expression can be obtained by simplifying the following Jacobian, corresponding to the coordinate transformation from $(\mathcal{M}_{c}^{z}, q, d_{L})$ to $(m_{1}, m_{2}, z)$:

\begin{equation} \label{eq:jacobian_matrix}
    J \left( \frac{\mathcal{M}_{c}^{z}, q, d_{L}}{m_{1}, m_{2}, z}\right) = 
    \begin{bmatrix}
  \frac{\partial \mathcal{M}_{c}^{z}}{\partial m_{1}} & 
    \frac{\partial \mathcal{M}_{c}^{z}}{\partial m_{2}} & 
    \frac{\partial \mathcal{M}_{c}^{z}}{\partial z} \\[2.5ex] 
  \frac{\partial q}{\partial m_{1}} & 
    \frac{\partial q}{\partial m_{2}} & 
    \frac{\partial q}{\partial z} \\[2.5ex]
  \frac{\partial d_{L}}{\partial m_{1}} & 
    \frac{\partial d_{L}}{\partial m_{2}} & 
    \frac{\partial d_{L}}{\partial z} 
\end{bmatrix}
\,.
\end{equation}
Using the definitions of luminosity distance (Eq.~\eqref{eq:luminosity_distance}) and comoving distance ($d_{c}=d_{L}/\left(1+z\right)$), Eq.~\eqref{jacobian} can be reexpressed as~\footnote{For a detailed derivation of the Jacobian, see Appendix~\ref{appendix:jacobian}.},
\begin{align} \label{eq:prior_transformation}
    p_{\rm PE}(m_{1}, m_{2}, z) \propto d_{L}^{2} (1+z)  & \left[d_{c}+\frac{c(1+z)}{H(z)} \right] \times \nonumber \\
    & \left[\frac{(m_{1} m_{2})^{3/5}}{m_{1}^{2} (m_{1}+m_{2})^{1/5}}\right]\,.
\end{align}
This expression is used as $p_{\rm PE} (\bm{\theta}_i)$ in Eq.~\eqref{likelihood} to construct the source-frame likelihood from the corresponding posterior.

In Eq.~\eqref{eq:bayesian1}, the likelihood $\mathcal{L}$ assumes that the individual events constitute an unbiased representation of the population. However, GW detectors are more sensitive to high-mass, nearby, and face-on sources. Consequently, the observed population does not truly follow the astrophysical population. So, the selection function $\beta (\bm{\Theta})$ has been included in Eq.~\eqref{eq:bayesian1} to mitigate this bias in estimating model parameters; it is defined as
\begin{equation} \label{eq:sel_func}
    \beta(\bm{\Theta}) = \int p_{\rm det} (\bm{\theta}_{j}) p(\bm{\theta}_{j} \mid \bm{\Theta}) d\bm{\theta}_{j}\,,
\end{equation}
where $p_{\rm det} (\bm{\theta}_{j})$ denotes the probability that an event with BNS parameters $\bm{\theta}_{j}$ is detected.
In this work, the selection term $\beta(\bm{\Theta})$ is evaluated via 
Monte Carlo integration over an ensemble of $N_{\rm inj}$ injected signals, which are drawn from a fiducial population and a cosmological model described by the hyperparameters $\bm{\Theta}_{0}$.
Subsequently, we determine the number $N_{\rm found}$ ($\leq N_{\rm inj}$) of the injected signals that cross the detection threshold~\citep{Tiwari:2017ndi, 2019RNAAS...3...66F}. The selection function is then estimated as follows:
\begin{equation} \label{eq:selection_effect_1}
    \hat{\beta} (\bm{\Theta}) = \frac{1}{N_{\rm inj}} \sum_{j=1}^{N_{\rm found}} \frac{p(\bm{\theta}_{j} \mid \bm{\Theta})}{p(\bm{\theta}_{j} \mid \bm{\Theta}_{0})} \,.
\end{equation}
Note that in the special case where the population and cosmological model is chosen to exactly match the fiducial model,  $\hat{\beta} (\bm{\Theta})$ simplifies to being just the fraction 
$N_{\rm found}/N_{\rm inj}$.
As with other works~\citep{Tiwari:2017ndi, 2019RNAAS...3...66F}, we use the point estimator $\hat{\beta}$ to estimate the hyperparameters described in Eq.~(\ref{eq:bayesian1}).
The details of the BNS population that is used to calculate the selection effect are given in Sec.~\ref{sec:simulation}.

\section{Models} \label{sec:models}

Our method has been applied to a set of simulated BNS events. We assume parameterized models of NS EoS, mass-redshift distribution of the sources, and cosmology to simulate them. We briefly discuss each model, along with the true values of the model parameters used to construct the mock GW catalog.

\subsection{EoS Model} \label{sec:eos_model}
The NS equation of state is constrained by combining insights from various observations and theoretical calculations across a range of densities~\citep[see, e.g.,][]{Raaijmakers:2019dks,Landry_2020PhRvD.101l3007L,Jiang:2019rcw,Traversi:2020aaa,Dietrich:2020efo,Miller:2021qha, Biswas:2020puz, Biswas:2021pvm, Biswas:2021paf, Tiwari:2023tkj, Biswas:2024hja}. GW detections from BNS mergers~\citep{TheLIGOScientific:2017qsa,Abbott:2018exr, LIGOScientific:2020aai}, mass and radius measurements from NICER for individual pulsars~\citep{Riley:2019yda,Miller:2019cac,Riley:2021pdl,Miller:2021qha,Choudhury:2024xbk}, and neutron skin thickness data~\citep{PREX:2021umo,CREX:2022kgg} from nuclear experiments all contribute valuable data about the EoS at different density ranges. Ab-initio nuclear theories, such as chiral effective field theory (see Refs.~\cite{Epelbaum:2008ga, Machleidt:2011zz,Hammer:2012id,Hebeler:2020ocj,Drischler:2021kxf}), provide constraints up to twice nuclear saturation density, while high-density EoS insights come from perturbative quantum chromodynamics~\citep{Komoltsev:2021jzg}. Together, these diverse datasets help refine our understanding of the EoS, particularly at high densities where they support heavy NS masses and favor a stiffer EoS.

To constrain the EoS of dense matter, one must aim to remove the systematic uncertainties of the modeling as much as possible to interpret the data properly. It is crucial to construct a consistent framework to connect the observational results from astrophysical detections to microscopic properties of dense matter combined with theoretical prediction of nuclear matter. For this purpose, we have used the hybrid+PP EoS parameterization that has been developed by \textcite{Biswas:2020puz} and used to provide multimessenger analyses of NS properties \citep{Biswas:2020xna,Biswas:2021yge,Biswas:2024hja,Char:2024kgo}. It has also been used to create a framework to infer Hubble constant directly from GW signals from BNS mergers in the future~\citep{Ghosh:2022muc}.

Here, we briefly review the hybrid nuclear+piecewise-polytrope (PP) EoS parameterization: Since the crust has minimal impact~\citep{Gamba:2019kwu, Biswas:2019ifs} on the macroscopic properties of NS, such as mass, radius, and tidal deformability, the conventional Baym-Pethick-Sutherland (BPS) EoS~\citep{1971ApJ...170..299B} is employed to model it within this framework.~\footnote{The effect of choosing other crustal EoSs (e.g., SLy~\citep{Douchin:2001sv}) will be explored elsewhere.} 
\begin{enumerate}
    \item The first component is the EoS around the nuclear saturation density ($\rho_0$), which can be well represented by the parabolic expansion of energy per nucleon $e(\rho, \delta)$ of asymmetric nuclear matter,
    \begin{equation}
        e(\rho,\delta) \approx  e_0(\rho) +  e_{\rm sym}(\rho)\delta^2,
    \end{equation}
where $e_0(\rho)$ is the energy of symmetric nuclear matter for which the number of protons is equal to the number of neutrons, $e_{\rm sym}$ is the energy of the asymmetric nuclear matter (commonly referred to as ``symmetry energy'' in literature), and $\delta \equiv \frac{\rho_p-\rho_n}{\rho_p+\rho_n}$ is the measure of asymmetry in the neutron number density $\rho_n$ and the proton number density $\rho_p$. Around $\rho_0$, both energies can be further expanded in a Taylor series,
\begin{eqnarray}
    e_0(\rho) &=&  e_0(\rho_0) + \frac{ K_0}{2}\chi^2 \label{eq:e0} +\,...,\\
    e_{\rm sym}(\rho) &=&  e_{\rm sym}(\rho_0) + L\chi + \frac{ K_{\rm sym}}{2}\chi^2 
    ..., \label{eq:esym}
    \end{eqnarray}
where $\chi \equiv (\rho-\rho_0)/3\rho_0 \ll 1$. 
We limit the Taylor expansion to the second order in $\chi$ since we only utilize this expansion up to $1.25 \rho_0$. The lowest order parameters are well constrained by experiments. Therefore, we fix them at their median values, such as $e_0(\rho_0) = -15.9$ MeV and $\rho_0 =0.16\,\rm{fm^{-3}}$. Consequently, the free parameters of this nuclear-physics-informed model include curvature of symmetric matter $K_{0}$, nuclear symmetry energy $e_{\rm sym}$, its slope $L$, and curvature of symmetric energy $K_{\rm sym}$. 
A survey based on $53$ experimental results performed in 2016~\citep{Oertel:2016bki} found values of $e_{\rm sym} (\rho_0) = 31.7 \pm 3.2$~MeV and $L = 58.7 \pm 28.1$ MeV. Using these values as priors, a Bayesian analysis performed by Biswas \textit{et al.}~\cite{Biswas:2020puz} that combined multiple astrophysical observations (GWs and X-rays) has already provided better constraints on these quantities: $e_{\rm sym} (\rho_0) = 32.0^{+3.05}_{-3.01}$ MeV and $L = 61.0^{+17.7}_{-16.0}$ MeV.
(See also Ref.~\cite{Tsang:2023vhh} on the implications of heavy ion collision constraints on NS EoS empirical parameters.)

    \item At higher densities, the empirical parameterization starts to break down. Following Ref.~\cite{Read:2008iy}, for densities above $1.25 \rho_0$, we adopt a three-piece piecewise-polytrope parameterization. This approach uses polytropic indices $\Gamma_{1}$, $\Gamma_{2}$, and $\Gamma_{3}$, with fixed transition densities at $10^{14.7}~\mathrm{g/cm^3}$ and $10^{15}~\mathrm{g/cm^3}$, respectively.    
\end{enumerate}

Finally, it is crucial to ensure that the parameterized EoS adheres to fundamental physical principles, particularly causality and the requirement for pressure to increase monotonically with density. Additionally, in compliance with the principles of special relativity, the speed of sound within the NS must not exceed the speed of light.

The injected NS EoS parameters for the simulated BNS events are listed in Table~\ref{tab:eos_pop_parameter} and the maximum mass corresponding to the injected EoS is $m_{\rm max}=2.25~M_{\odot}$.
Instead of explicitly using all these NS EoS parameters, we will generally refer to them as $\bm{\Theta}_{\mathcal{E}}\equiv \{ K_{0}, e_{\rm sym}, L, K_{\rm sym}, \Gamma_{1}, \Gamma_{2}, \Gamma_{3} \}$.

\subsection{Population and Cosmology} \label{sec:pop_cosmo}

In this study, we assume that all NSs originate from a common mass distribution  $p(m \mid \bm{\Theta}_{m})$ and form BNSs with random pairing,
\begin{align}
    p (m_{1}, m_{2} \mid \bm{\Theta}_{m}) \propto p(m_{1} \mid \bm{\Theta}_{m}) p(m_{2} \mid \bm{\Theta}_{m}) \mathcal{H}(m_{2} > m_{1}) \,,
\end{align}
where $\mathcal{H}$ is the Heaviside step function enforcing that the primary mass $m_{1}$ be greater than the secondary mass $m_{2}$. There may be a possibility that each companion of a BNS follows a different mass distribution~\citep{Golomb:2021tll} due to different stellar evolutionary processes prior to collapse~\citep{Farrow:2019xnc}. However, we restrict the common mass distribution for both components of the binaries in this work for simplicity and the ability to capture the key features of the observed NS mass distribution.

In this work, we investigate two observationally motivated mass models for NSs between minimum mas $m_{\rm min}$ and maximum mass $m_{\rm max}$, to study their impact on the inference of other hyperparameters.~\footnote{Note that here $m_{\rm max}$ is being treated as a variable whose value is to be determined from observations. This should be contrasted with the usage where $m_{\rm max}$ defines a specific value of a simulated signal injection, as done previously. This dichotomy in usage is present for other parameters as well. The context in which they are used will resolve whether we are employing them as a source parameter to be inferred from observations or a specific injection value.}
To ensure that the mass distribution has vanishing support 
outside the interval $[m_{\rm min}, m_{\rm max}]$, we multiply the probability density function of mass by $\mathcal{H} (m_{\rm min}, m_{\rm max})=\mathcal{H}(m > m_{\rm min}) \mathcal{H}(m < m_{\rm max})$. We set $m_{\rm min} = 1 M_{\odot}$, which is consistent with the predicted lower bound of NS mass from plausible supernova formation channels~\citep{2012ApJ...749...91F, Woosley:2020mze}. The maximum mass, $m_{\rm max}=2.25 M_{\odot}$, is supported by the injected NS EoS parameters noted in Sec.~\ref{sec:eos_model}.

\begin{enumerate}
    
    \item Gaussian distribution: This mass distribution is primarily motivated by the Gaussian fit to the mass distribution of Galactic NSs~\citep{Ozel:2012ax, Kiziltan:2013oja, Ozel:2016oaf}:
    \begin{align} \label{eq:mass_gaussian}
        p_{\mathcal{N}}(m \mid m_{\rm min}, m_{\rm max}, \mu, \sigma) \propto & \>\mathcal{N}(m \mid \mu, \sigma) \times \nonumber\\
        & \mathcal{H} (m_{\rm min}, m_{\rm max})\,.
    \end{align}
    The Gaussian distribution characterized by mean $\mu$ and standard deviation $\sigma$ is defined in the range $\left[m_{\rm min}, m_{\rm max}\right]$. So, the mass hyperparameters corresponding to the Gaussian distribution are $\bm{\Theta}_{m}=\left\{m_{\rm min}, m_{\rm max}, \mu, \sigma\right\}$.
    Following Ref.~\cite{Ozel:2016oaf}, we set $\mu=1.33~M_{\odot}$ and $\sigma=0.09~M_{\odot}$ as the injected values of the parameters of the Gaussian mass distribution.

    \item Double Gaussian distribution: In this case, the NS mass distribution is taken to follow a double Gaussian distribution, 
    which is the same as the following two-component Gaussian distribution~\citep{Alsing:2017bbc, Shao:2020bzt}:
    \begin{align} \label{eq:mass_double_gaussian}
        p_{\mathcal{NN}} (m \mid \bm{\Theta}_{m}) & \propto  \Big[w\mathcal{N}(m \mid \mu_{1}, \sigma_{1}) + \nonumber \\
        & (1-w)\mathcal{N}(m \mid \mu_{2}, \sigma_{2})\Big] \mathcal{H} (m_{\rm min}, m_{\rm max})\,.
    \end{align}
    Here, $\bm{\Theta}_{m}=\left\{m_{\rm min}, m_{\rm max}, \mu_{1}, \sigma_{1}, \mu_{2}, \sigma_{2}, w\right\}$ corresponds to the mass-distribution hyperparameters, consisting of the mean $\mu_{1}$ ($\mu_{2}$), the standard deviation $\sigma_{1}$ ($\sigma_{2}$), and the relative weight of two components $w$. We consider the mass distribution of NSs (in our simulated BNSs) to be the double Gaussian with $\mu_{1}=1.34~M_{\odot}$, $\sigma_{1}=0.07~M_{\odot}$, $\mu_{2}=1.8~M_{\odot}$, $\sigma_{2}=0.21~M_{\odot}$, and $w=0.65$. This distribution is consistent with the NS mass distribution from pulsar mass measurements~\citep{Alsing:2017bbc, Shao:2020bzt}.

\end{enumerate}

Furthermore, we assume that the BNS mass distribution does not evolve with cosmic time (i.e., it is independent of redshift). This is a reasonable assumption for the low-redshift BNSs, which will be observed by the current-generation ground-based detectors.
However, one can relax this assumption for the next-generation detectors, which will have larger horizon redshifts for BNS mergers and could probe whether the NS mass distribution varies across cosmic time.

\begingroup
\begin{table*}
\begin{center}
\begin{tabular}{ |c|c|c|c|c| } 
  \hline
  Model & Parameters & Units & True Values & Priors \\ [0.5ex] 
  \hline
  \multirow{7}{3.2em}{EoS} & $K_{0}$ & MeV & $239.7$ & $U(129,350)$\\ 
                         & $e_{\rm sym}$ & MeV &  $32.5$ & $U(21,43)$\\ 
                         & $L$ & MeV &  $69$ & $U(8,145)$\\
                         & $K_{\rm sym}$ & MeV &  $-174.6$ & $U(-560,251)$\\
                         & $\Gamma_{1}$ &  - & $3$ & $U(0.2,7.8)$\\
                         & $\Gamma_{2}$ &  - & $4$ & $U(1.2,6.1)$\\
                         & $\Gamma_{3}$ &  - & $3.7$ & $U(0.2,8)$\\
  \hline
  Redshift & $\gamma$ & - &  $0$ & $U(-20, 20)$\\
  \hline
  \multirow{2}{5em}{Cosmology} & $H_{0}$ & $\kmsMpc$ &  $70$ & $U(10, 200)$\\
                               & $\Omega_{m}$ & $-$ &  $0.3$ & $U(0, 1)$\\
  \hline
\end{tabular}
\caption{\label{tab:eos_pop_parameter}Summary of hyperparameters for the EoS, redshift, and cosmological model used in this work. The details of hyperparameters for mass models are provided in Table~\ref{tab:mass_model_parameters}.
By ``True Values," we mean the parameter values used for simulating the BNS signals and their populations.}
\end{center}
\end{table*}
\endgroup

\begingroup
\begin{table*}
\begin{center}
\begin{tabular}{ |c|c|c|c|c| } 
  \hline
  Mass Model & Parameters & Units & True Values & Priors \\ [0.5ex] 
  \hline
  \multirow{1}{5em}{Gaussian} & $\mu$ & $M_{\odot}$ &  $1.33$& $U(1,m_{\rm max})$\\ 
                          & $\sigma$ & $M_{\odot}$  &  $0.09$ & $U(0.005, 0.5)$\\
  \hline
  \multirow{4}{5em}{\begin{tabular}{@{}c@{}}Double \\ Gaussian\end{tabular}} & $\mu_{1}$ & $M_{\odot}$ &  $1.34$& $U(1,2)$\\ 
                          & $\sigma_{1}$ & $M_{\odot}$  &  $0.07$ & $U(0.005, 0.5)$\\
                          & $\mu_{2}$ & $M_{\odot}$ &  $1.8$& $U(\mu_{1},m_{\rm max})$\\ 
                          & $\sigma_{2}$ & $M_{\odot}$  &  $0.21$ & $U(0.005, 0.5)$\\
                          & $w$ & $-$  &  $0.65$ & $U(0, 1)$\\
  \hline
\end{tabular}
\caption{\label{tab:mass_model_parameters}Summary of hyperparameters for $2$ different mass models used in this work. $m_{\rm max}$ is chosen between $[1, 3]~M_{\odot}$ when EoS is not considered; otherwise $m_{\rm max}$ is determined from the NS EoS. All mass models are defined within the fixed mass range $[m_{\rm min}, m_{\rm max}]=[1, 2.25]~M_{\odot}$.}
\end{center}
\end{table*}
\endgroup

The redshift distribution of BNS mergers can be written as
\begin{equation} \label{eq:redshift_model}
    p(z) \propto \frac{dV_{c}}{dz} \frac{\mathcal{R}(z)}{1+z} \\,
\end{equation}
where $V_{c}$ is the comoving volume and $\mathcal{R}(z)$ is the source-frame BNS merger rate as a function of redshift. 
The term $(1+z)$ in the denominator of Eq.~\eqref{eq:redshift_model} is used to convert the source-frame time to the detector-frame time.
Considering merger rate $\mathcal{R}(z)$ to be a power law, i.e., $\mathcal{R} (z) \propto (1+z)^{\gamma}$, the corresponding redshift distribution~\citep{KAGRA:2021duu} of BNSs is 
\begin{equation} 
    p(z \mid \gamma) \propto \frac{dV_{c}}{dz} (1+z)^{\gamma -1} \,.
\end{equation}
In our work, we set the merger-rate index $\gamma=0$ for creating the mock BNS catalog, 
implying that the BNS merger rate is uniform in comoving volume and source-frame time, which is consistent with the choices adopted by LVK~\citep{KAGRA:2013rdx, LIGOScientific:2020kqk, KAGRA:2021duu}.
We assume flat $\Lambda$CDM cosmology to convert the redshift to luminosity distance by using the following relation:

\begin{equation} \label{eq:luminosity_distance}
    d_{L} (z) = \frac{c(1+z)}{H_{0}} \int_{0}^{z} \frac{dz^{\prime}}{\sqrt{\Omega_{m}(1+z^{\prime})^{3}+(1-\Omega_{m})}}\,,
\end{equation}
where $c$ and $\Omega_{m}$ correspond to the speed of light and matter density, respectively.
For our simulations, we employ $\Lambda$CDM cosmology with $H_{0}=70~\kmsMpc$ and $\Omega_{m}=0.3$ as the true cosmology.~\footnote{Later discussion shows that with our samples of BNSs, it is not possible to constrain $\gamma$ or the cosmological parameters, apart from $H_0$, with current-generation detectors. Even so, we include them in our study for the completeness of the parameters that have a bearing on the Bayesian parameter estimation results reported here.}

\begin{figure*} 
    \centering
    \includegraphics[scale=0.78]{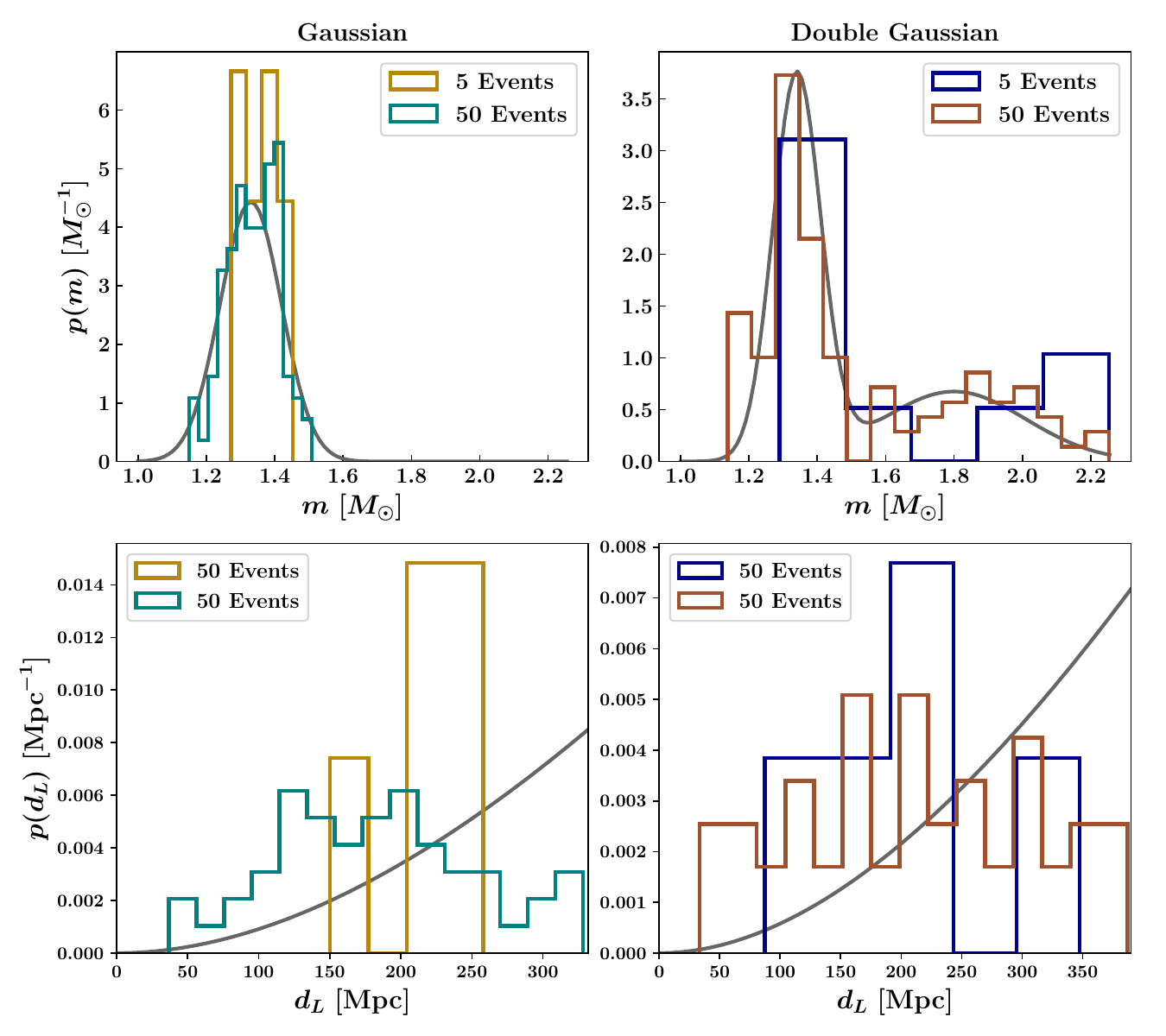}
    \caption{\emph{Top Row:} The left and right panels show the injected and detected component mass distributions of NSs for the Gaussian and double Gaussian mass models, respectively. The black solid lines indicate the true mass distributions (see Table~\ref{tab:mass_model_parameters}), and histograms correspond to the source-frame masses of $5$ and $50$ NSs detected with SNR $\geq 20$.
    \emph{Bottom Row:} The left and right panels display the luminosity distance distributions of the detected events for the same mass models. The black lines represent the underlying true distributions, while the histograms show results for $5$ and $50$ detected NSs with SNR $\geq 20$.}
    \label{fig:mass_hist}
\end{figure*}

\section{Simulation} \label{sec:simulation}

\begin{figure*} 
    \centering
    \includegraphics[scale=0.5]{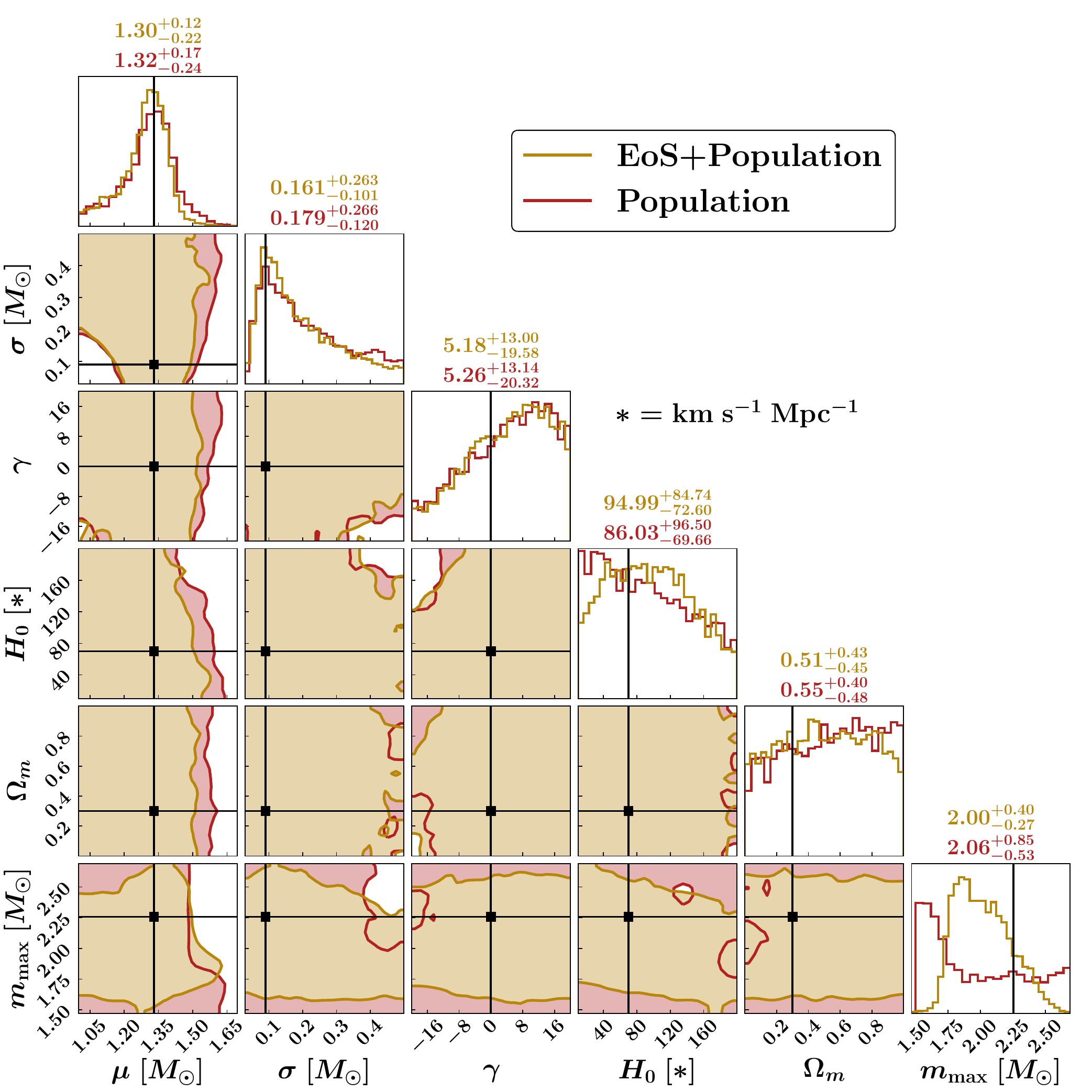}
    \caption{Comparison of the inferred population and cosmological parameters from $5$ events, based on the Gaussian mass distribution, detected by the LIGO-Virgo detectors. The label `Population' refers to the Bayesian analysis considering only mass and redshift information ($m_{1}, m_{2}, z$), while the other label `EoS+Population' also includes tidal parameters ($m_{1}, m_{2}, z, \Lambda_{1}, \Lambda_{2}$). The black solid lines indicate the injected values of the corresponding parameters. The $90\%$ credible intervals are shown for each of the respective marginalized $1$D posteriors.}
    \label{fig:pop_parameters_compare_5events_gaussian}
\end{figure*}

\begin{figure*} 
    \centering
    \includegraphics[scale=0.5]{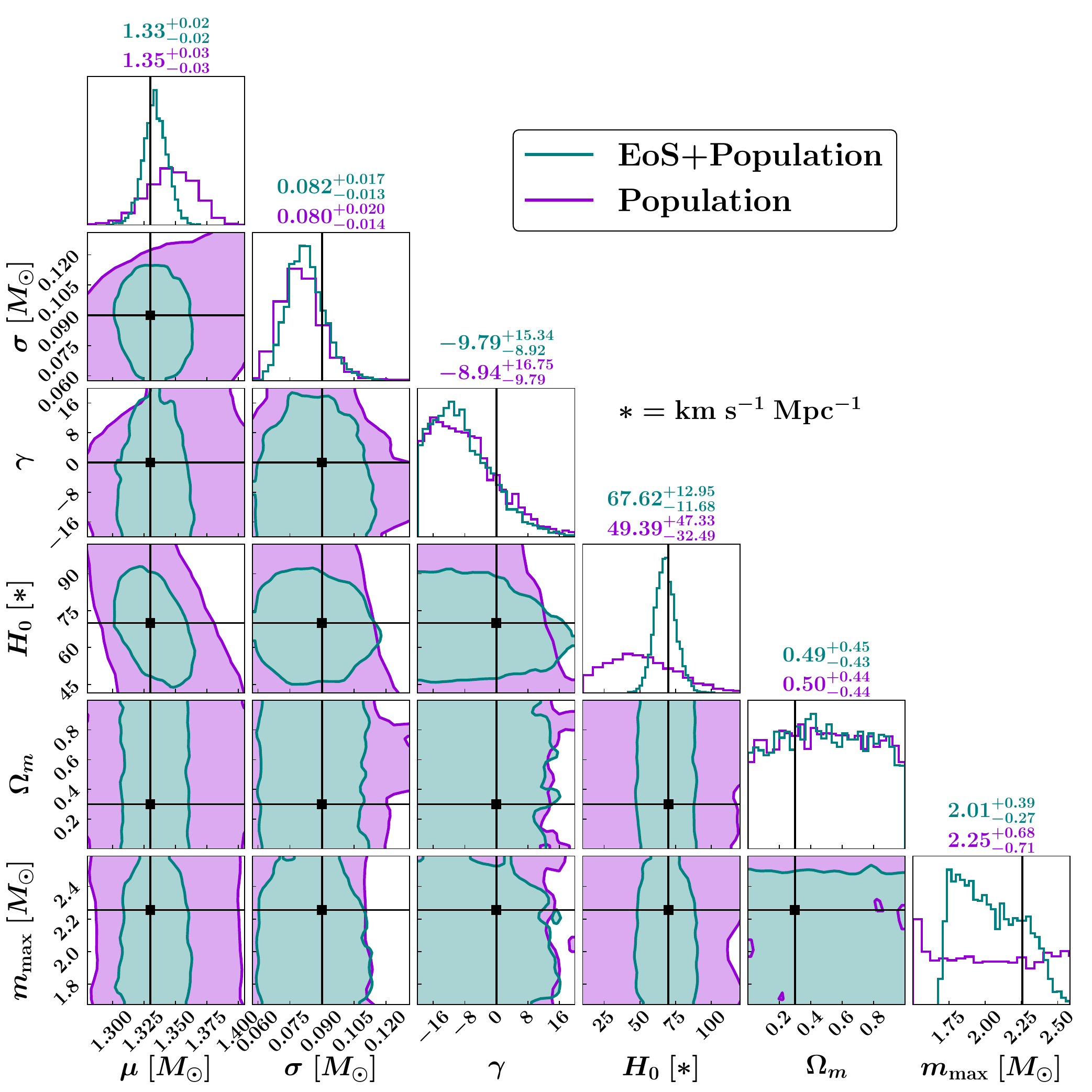}
    \caption{Same as Fig.~\ref{fig:pop_parameters_compare_5events_gaussian}, but using $50$ events.}
    \label{fig:pop_parameters_compare_50events_gaussian}
\end{figure*}

\begin{figure*} 
    \centering
    \includegraphics[scale=0.34]{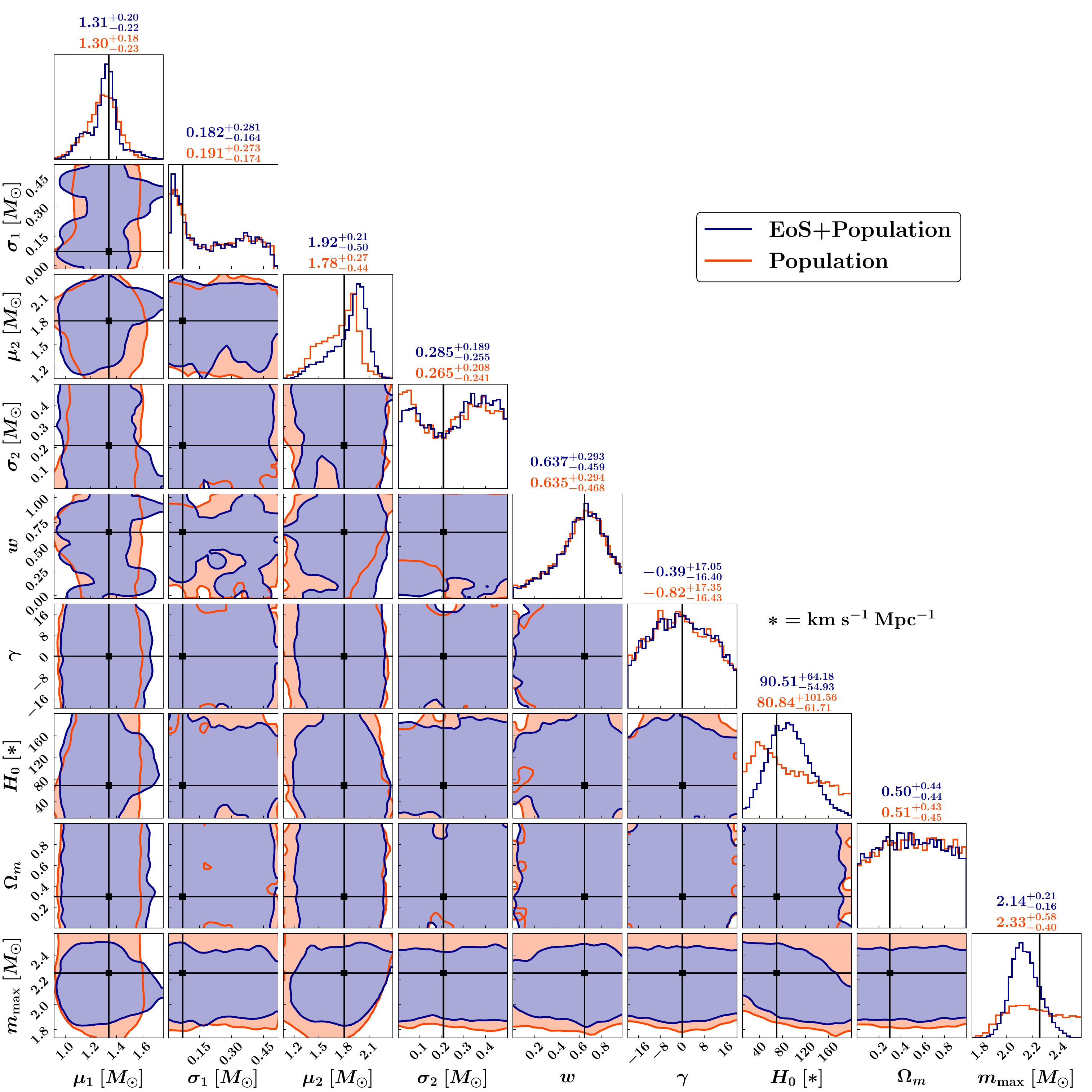}
    \caption{Comparison of the inferred population and cosmological parameters from $5$ GW events, following the double Gaussian mass distribution, detected by the LIGO-Virgo detectors. The black solid lines show the injected values of the corresponding parameters. The $90\%$ credible intervals are also mentioned for each of the respective marginalized $1$D posteriors.}
    \label{fig:pop_parameters_compare_5events_double_gaussian}
\end{figure*}

\begin{figure*} 
    \centering
    \includegraphics[scale=0.34]{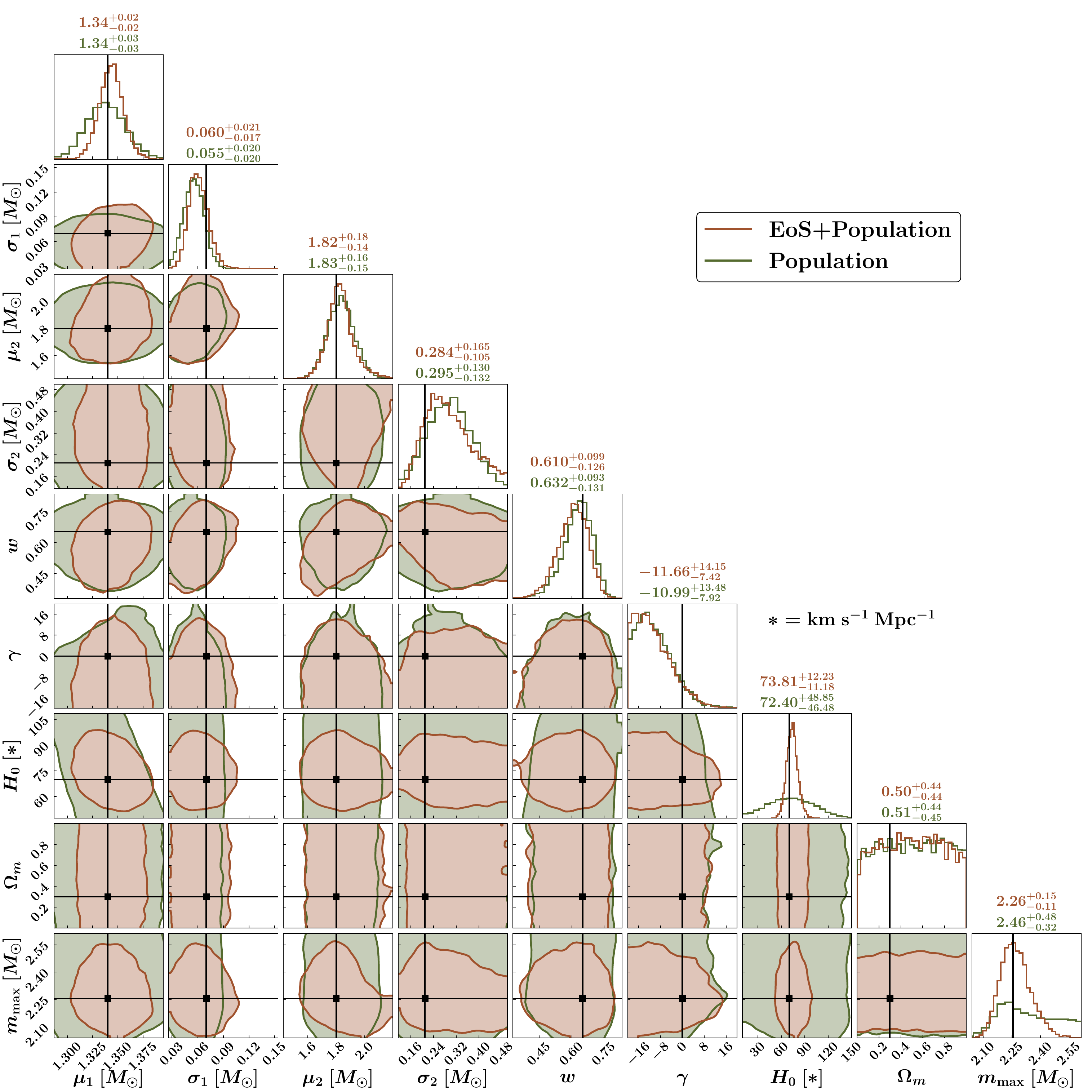}
    \caption{Same as Fig.~\ref{fig:pop_parameters_compare_5events_double_gaussian}, but using $50$ events.}
    \label{fig:pop_parameters_compare_50events_double_gaussian}
\end{figure*}

\begin{figure*}
    \centering
    \includegraphics[scale=0.55]{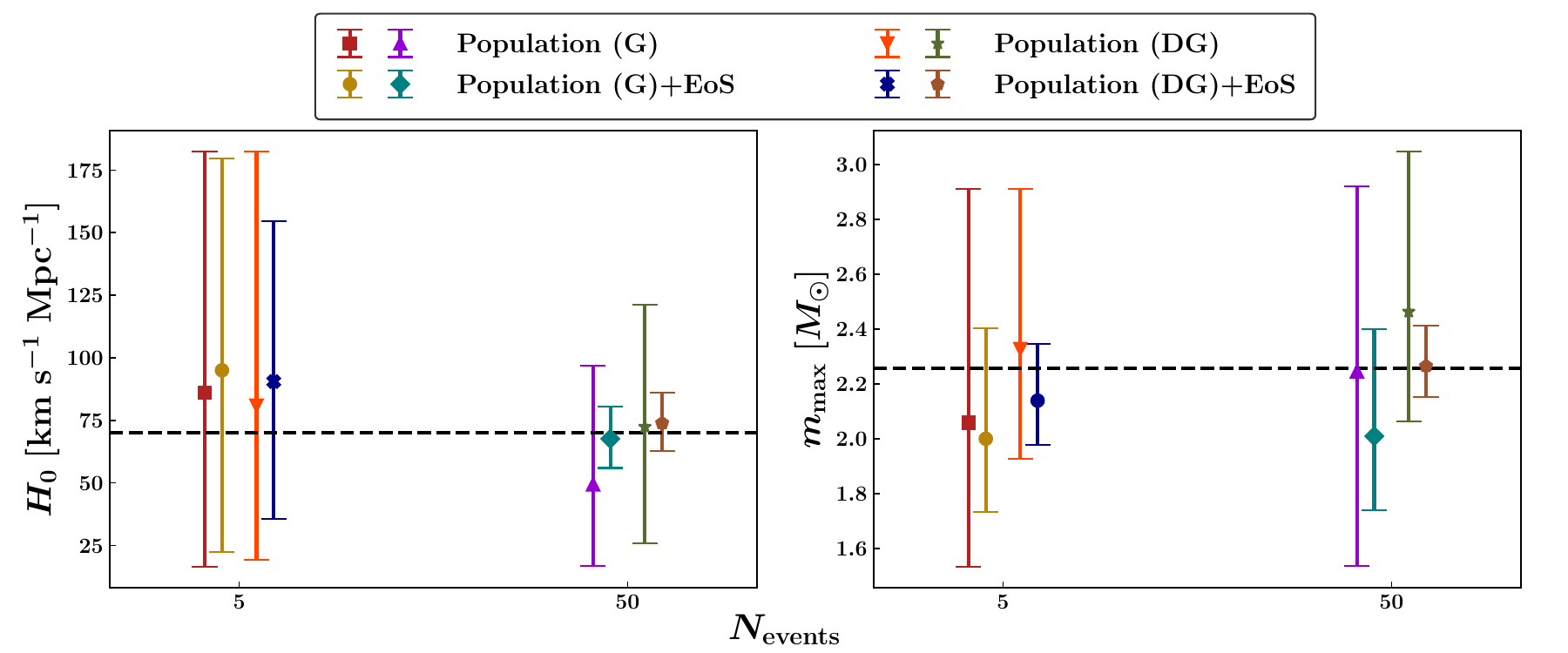}
    \caption{Comparison of the uncertainty in measuring $H_{0}$ and $m_{\rm max}$ from $5$ and $50$ GW events, following the Gaussian (G) and double Gaussian (DG) mass distributions. The $x$-axis is not to scale.}
    \label{fig:H0error_compare}
\end{figure*}

We demonstrate the efficacy of our method with a set of simulated BNS signals in a three-detector network, comprising two Advanced LIGO detectors (in Hanford and Livingston)~\citep{LIGOScientific:2014pky} and the Advanced Virgo detector~\citep{VIRGO:2014yos}, operating at their respective O5 design sensitivities~\citep{KAGRA:2013rdx}.~\footnote{\url{https://dcc.ligo.org/LIGO-T2000012/public}} 
We construct a mock GW catalog comprising $50$ events uniformly distributed across the sky.
The merger rate is uniform in comoving volume and source-frame time, while component masses are drawn from Gaussian and double Gaussian distributions between the minimum mass $m_{\rm min}=1~M_{\odot}$ and the maximum mass $m_{\rm max}=2.25~M_{\odot}$, as specified in Sec.~\ref{sec:pop_cosmo}.
We utilize only those GW sources that are detected with network signal-to-noise (SNR) $\rho \geq 20$ (see Fig.~\ref{fig:mass_hist}). 
This threshold is chosen because events with $\rho \geq 20$ provide significant information about the EoS, which is essential for constraining the Hubble constant~\citep{Golomb:2021tll}.
Applying this SNR threshold, we find that the detected simulations are distributed 
within $\sim 330$ Mpc for the Gaussian mass distribution and $\sim 390$ Mpc for the double Gaussian mass distribution, which are the typical ranges for BNS detections in the O5 era~\citep{KAGRA:2013rdx}.
The models for the NS EoS, BNS population, and cosmology are mentioned in Sec.~\ref{sec:models}, along with the corresponding injected values, which are also summarized in Table~\ref{tab:eos_pop_parameter}. 
We assume the NSs to be slowly rotating, with the dimensionless spin parameter restricted to $\chi \leq 0.05$~\citep{LIGOScientific:2018hze}, consistent with spin measurements of NSs in Galactic binary systems~\citep{Tauris:2017omb}.
This is a reasonable approximation, as by the time a BNS signal enters the LIGO-Virgo frequency band, the NSs are generally old, and much of their spin has likely decayed during their orbital evolution~\citep{Shapiro:1983du}. We simulate the BNS signal using the waveform model, \textsc{IMRPhenomPv2\_NRTidal}~\citep{Dietrich:2019kaq}, injected in the simulated colored Gaussian noise corresponding to the three detectors.

We perform Bayesian parameter estimation of the simulated GW strain data between $20$~Hz and $2048$~Hz using the nested sampler \textsc{dynesty}~\citep{2020MNRAS.493.3132S} implemented in \textsc{bilby\_pipe}~\citep{Ashton_2019}. We use the same waveform model employed for injection to obtain the posterior probability distribution of BNS parameters. 
We perform parameter estimation over all BNS parameters (see Table 2 in Ref.~\cite{Ashton_2019} for the complete list), as is done for real GW data, to account for realistic uncertainties in the parameters considered in this study, such as the mass pair, tidal deformabilities, and luminosity distance.
We consider a uniform prior over the observed chirp mass and mass ratio, such that the detector-frame component masses lie within the range $[1-4]~M_{\odot}$.
For the luminosity distance, we employ a cosmology-independent $d_{L}^{2}$ prior between $10$ Mpc and $1$ Gpc. 
The priors on the tidal deformabilities of each binary component are taken to be uniform between $0$ and $5000$.
The priors on the remaining parameters are the default priors implemented in \textsc{bilby} (refer to Table $2$ in~\citep{Ashton_2019} for the default priors).
In Appendix~\ref{appendix:m_lambda_posterior}, Fig.~\ref{fig:m_lambda_corner} presents the marginalized joint posteriors of the detected parameters of a BNS event. These parameters are involved in inferring population, NS EoS, and cosmology. 

Once we obtain the posterior samples of all detected individual events, we combine them to constrain the NS EoS, mass-redshift distribution, and cosmological parameters, following the formalism outlined in Sec.~\ref{sec:method}. We consider uniform priors for all the model hyperparameters to be inferred in this study. 
The priors of the hyperparameters are listed in Tables~\ref{tab:eos_pop_parameter}~\footnote{To remain conservative, 
we take our priors on nuclear parameters such as $L$ and $e_{\rm sym}$ to be broader than what was mentioned earlier in Sec.~\ref{sec:eos_model}.} and \ref{tab:mass_model_parameters}. 
To compute the likelihood, we convert the detector-frame chirp mass and mass ratio, inferred from GW strain data assuming uniform priors on both, to source-frame component masses, as required for model parameter inference in Eq.~\eqref{eq:bayesian1}. Moreover, from those priors we need to deduce the respective priors on the component masses (see Eq.~\eqref{eq:prior_transformation}).

For an unbiased estimation of hyperparameters in Eq.~\eqref{eq:bayesian1}, we calculate the selection function $\beta$ by taking the NSs to follow a uniform mass distribution in the source frame, with masses between $1$ and $3~ M_{\odot}$. Although the injected maximum mass of the simulated events is $2.25~M_{\odot}$, we use $3~M_{\odot}$ as the maximum mass for 
computing the selection function because varying NS EoS parameters during their inference also implies that the maximum mass will be varied 
(see the 1D marginalized effective prior of $m_{\rm max}$ in Fig.~\ref{fig:EoS_prior_corner}). 
For each BNS signal, we compute the SNR of the simulated BNS signals in the O5 noise realization and thereby determine the selection function following Eq.~\eqref{eq:selection_effect_1}. 
We do not consider the impact of the tidal effect in calculating the selection function because the tidal effect starts appearing at the $5$ post-Newtonian order and has no significant contribution to the amplitude of the GW signal. We have used the nested sampler \textsc{Pymultinest}~\citep{Buchner:2014nha} for performing the Bayesian inference of Eq.~\eqref{eq:bayesian1}.

\begingroup
\begin{table*}
\begin{center}
\begin{tabular}{| c | c@{\hskip 1.1em} | c@{\hskip 1.1em} | c@{\hskip 1.1em} | c@{\hskip 1.1em}| c@{\hskip 1.1em} |}
\hline
\multirow{2}{*}{\centering Mass Model} & \multirow{2}{*}{\centering Parameters} & \multicolumn{2}{c|}{$5$ Events} & \multicolumn{2}{c|}{$50$ Events} \\
\cline{3-6}
& & Population & EoS+Population & Population & EoS+Population \\
\hline
\multirow{6}{5em}{\centering Gaussian}  & $\mu$ & $1.32^{+0.17}_{-0.24}$ & $1.30^{+0.12}_{-0.22}$ & $1.35^{+0.03}_{-0.03}$ & $1.33^{+0.02}_{-0.02}$ \\
& $\sigma$ & $0.179^{+0.266}_{-0.120}$ & $0.161^{+0.263}_{-0.101}$ & $0.080^{+0.020}_{-0.014}$ & $0.082^{+0.017}_{-0.013}$ \\
& $\gamma$ & $5.26^{+13.14}_{-20.32}$ & $5.18^{+13.00}_{-19.58}$ & $-8.94^{+16.75}_{-9.79}$ & $-9.79^{+15.34}_{-8.92}$ \\
& $H_{0}$ & $86.03^{+96.50}_{-69.66}$ & $94.99^{+84.74}_{-72.60}$ & $49.39^{+47.33}_{-32.49}$ & $67.62^{+12.95}_{-11.68}$ \\
& $\Omega_{m}$ & $0.55^{+0.40}_{-0.48}$ & $0.51^{+0.43}_{-0.45}$ & $0.50^{+0.44}_{-0.44}$ & $0.49^{+0.45}_{-0.43}$ \\
& $m_{\rm max}$ & $2.06^{+0.85}_{-0.53}$ & $2.00^{+0.40}_{-0.27}$ & $2.25^{+0.68}_{-0.71}$ & $2.01^{+0.39}_{-0.27}$ \\
\hline

\multirow{7}{5em}{\centering Double Gaussian}  & $\mu_{1}$ & $1.30^{+0.18}_{-0.23}$ & $1.31^{+0.20}_{-0.22}$ & $1.34^{+0.03}_{-0.03}$ & $1.34^{+0.02}_{-0.02}$ \\
& $\sigma_{1}$ & $0.191^{+0.273}_{-0.174}$ & $0.182^{+0.281}_{-0.164}$ & $0.055^{+0.020}_{-0.020}$ & $0.060^{+0.021}_{-0.017}$ \\
& $\mu_{2}$ & $1.78^{+0.27}_{-0.44}$ & $1.92^{+0.21}_{-0.50}$ & $1.83^{+0.16}_{-0.15}$ & $1.82^{+0.18}_{-0.14}$ \\
& $\sigma_{2}$ & $0.265^{+0.208}_{-0.241}$ & $0.285^{+0.189}_{-0.255}$ & $0.295^{+0.130}_{-0.132}$ & $0.284^{+0.165}_{-0.105}$ \\
& $w$ & $0.635^{+0.294}_{-0.468}$ & $0.637^{+0.293}_{-0.459}$ & $0.632^{+0.093}_{-0.131}$ & $0.610^{+0.099}_{-0.126}$ \\
& $\gamma$ & $-0.82^{+17.35}_{-16.43}$ & $-0.39^{+17.05}_{-16.40}$ & $-10.99^{+13.48}_{-7.92}$ & $-11.66^{+14.15}_{-7.42}$ \\
& $H_{0}$ & $80.84^{+101.56}_{-61.71}$ & $90.51^{+64.18}_{-54.93}$ & $72.40^{+48.85}_{-46.48}$ & $73.81^{+12.23}_{-11.18}$ \\
& $\Omega_{m}$ & $0.51^{+0.43}_{-0.45}$ & $0.50^{+0.44}_{-0.44}$ & $0.51^{+0.44}_{-0.45}$ & $0.50^{+0.44}_{-0.44}$ \\
& $m_{\rm max}$ & $2.33^{+0.58}_{-0.40}$ & $2.14^{+0.21}_{-0.16}$ & $2.46^{+0.48}_{-0.32}$ & $2.26^{+0.15}_{-0.11}$  \\
\hline

\end{tabular}
\caption{\label{tab:pop_posterior} Comparison of the $90\%$ credible interval corresponding to the uncertainty in measuring the population parameters with and without EoS parameters from GW events of $5$ and $50$, following Gaussian and double Gaussian mass distributions.}
\end{center}
\end{table*}
\endgroup

\section{Results} \label{sec:results}

In this study, we explore how well the NS population and the cosmological parameters can be inferred from GW observations of BNSs, especially when the NS EoS parameters are also measured. 
The primary objective of this study is to explore the limits of $H_0$ estimation during the O5 era with those dark binaries.
To pursue this, we simulate two sets of BNSs  -- one with $5$ events and the other with $50$ events (including those $5$ events), as detected during the O5 era of the LIGO and Virgo detectors, with the network SNR $\geq 20$.
The choice of $5$ BNSs is a realistic detection scenario for the O5 era and serves to investigate the possible recovery of population, cosmology, and NS EoS.
These $5$ simulated BNSs, which follow the Gaussian mass distribution in one experiment and the double Gaussian one in another, would be consistent with a merger rate of $R_{0}\sim 1200~{\rm Gpc^{-3}yr^{-1}}$ and $R_{0}\sim 700~{\rm Gpc^{-3}yr^{-1}}$ for those two experiments, respectively -- assuming that the O5 run will last $2$ years~\citep{KAGRA:2021duu}.
On the other hand, the choice of $50$ events, which is relatively high compared to the current best estimate of the merger rate, allows us to assess the statistical robustness of our methodology and investigate any potential systematic effects that dominate our methodology.
Note that in neither of the mass models, the impact of different redshift evolution models in constraining hyperparameters is explored. This is because all our simulated GW events are confined to the local Universe, i.e., within $\sim 390$~Mpc.

\subsection{Inference of Population and Cosmological Parameters} \label{subsec:result_1}

The methodology described in Sec.~\ref{sec:method} is applied to 
the simulated events described previously.
Initially, we study the influence of the NS EoS on inferring the population and the Hubble constant. 
The Bayesian analysis is carried out both without and with accounting for the NS EoS. 
The former analysis considers only mass and redshift information ($m_{1}, m_{2}, z$), while the latter also includes tidal parameters 
($\Lambda_{1}, \Lambda_{2}$).
In our various parameter estimation figures, these analyses are referred to as `Population' and `EoS+Population,' respectively.
Figures~\ref{fig:pop_parameters_compare_5events_gaussian} and \ref{fig:pop_parameters_compare_50events_gaussian} present the estimation of the population and the cosmological parameters for $5$ and $50$ events, respectively, with BNSs following the Gaussian mass distribution. In each figure, the population and the cosmological parameters are inferred using both approaches: `Population' (considering only mass and redshift information) and `EoS+Population' (including tidal parameters).
Similar comparisons for BNSs with the double Gaussian mass distribution are shown in Figures~\ref{fig:pop_parameters_compare_5events_double_gaussian} and \ref{fig:pop_parameters_compare_50events_double_gaussian}.
The uncertainties corresponding to the $90\%$ credible intervals of all the population and the cosmological parameters are summarized in Table~\ref{tab:pop_posterior}.

These results show that incorporating the NS EoS into the spectral siren method significantly improves the constraint on $H_{0}$ and $m_{\rm max}$. For a given NS EoS, the NS of (source-frame) mass $m$ uniquely determines its tidal deformability $\Lambda$. This $m-\Lambda$ relation is determined by solving the Tolman-Oppenheimer-Volkoff (TOV) equations~\citep{PhysRev.55.364, PhysRev.55.374} for a specific NS EoS, as elaborated in Appendix~\ref{appendix:tov}.~\footnote{In case no such $m-\Lambda$ correlations were accounted for, 
the posteriors of the BNS parameters from GW data would turn out to be as shown in Fig.~\ref{fig:m_lambda_corner} of Appendix~\ref{appendix:m_lambda_posterior}.}
Once $\Lambda$ is measured from a BNS signal, it can be used to deduce the source-frame mass by utilizing this $m-\Lambda$ relation. Comparing that source-frame mass with the detected mass $m^{z}$ provides the redshift of the source, following the relation $m^{z} = m(1+z)$. 
When combined with the distance posteriors of the GW observations, these inferred redshifts are able to constrain $H_0$.
In reality, there is uncertainty in the NS EoS parameters, which propagates into the inferred redshift and, consequently, into the estimation of $H_0$.

Additionally, the NS EoS helps constrain the maximum mass from GW observations since $m_{\rm max}$ is a derived parameter from the NS EoS rather than a free parameter.
Hence, incorporating the EoS into the analysis naturally tightens its constraint. 
While employing a population model alone provides some information on $m_{\rm max}$, the inclusion of the NS EoS leads to a significantly stronger constraint on that parameter.
Whereas $m_{\rm max}$ is primarily a population parameter, its precise inference is strongly correlated with the constraints on the NS EoS. Therefore, we address the improved $m_{\rm max}$ constraints along with the underlying reasons in Sec.~\ref{subsec:result_2}.

Importantly, the precision of $m_{\rm max}$ also influences the inference of $H_0$~\citep{Mastrogiovanni:2021wsd}. 
For lower sample values of $H_0$, GW sources at any given luminosity distance will be deduced to be at lower redshifts, thereby making their source-frame masses appear larger. If these inferred masses approach or exceed $m_{\rm max}$, they become incompatible with the assumed population model. Conversely, higher sampled $H_0$ values lead to lower inferred source-frame masses, which also need to align with the mass distribution governed by $m_{\rm max}$. Additionally, $m_{\rm max}$ controls the fraction of detected events at the high-mass end. A lack of such expected events at higher mass can be compensated by lowering either $m_{\rm max}$ or $H_0$. Therefore, the interplay between $H_0$ and $m_{\rm max}$ is crucial for matching the observed mass distribution to the population model.
In summary, incorporating the NS EoS into the spectral siren method ensures consistency with the expected $m$–$\Lambda$ correlation for NSs. This EoS-informed approach enhances the precision of the inferred model parameters.
However, a study of systematic errors and their sources in the estimation of these parameters has not been explored in this work and can be pursued in future investigations.

Notably, $H_{0}$ has been inferred with large errors: $H_{0} = 86.03^{+96.50}_{-69.66}$~\kmsMpc and $80.84^{+101.56}_{-61.71}$~\kmsMpc from $5$ GW events with the Gaussian and double Gaussian mass distributions, respectively, when the NS EoS is not utilized.
However, $50$ events following both the mass distributions can infer $H_{0}$ without considering the NS EoS, yielding $H_{0} = 49.39^{+47.33}_{-32.49}$~\kmsMpc and $H_{0} = 72.40^{+48.85}_{-46.48}$~\kmsMpc, for the Gaussian and double Gaussian mass distributions, respectively. 
Nevertheless, the precision of the $H_{0}$ measurement is improved by a factor of $\gtrsim 3$ when the NS EoS is utilized, as demonstrated with $50$ simulated events in Fig.~\ref{fig:pop_parameters_compare_50events_gaussian} and Fig.~\ref{fig:pop_parameters_compare_50events_double_gaussian}, corresponding to the Gaussian and double Gaussian mass distributions, respectively.
In comparison, the inference of the other cosmological parameter, $\Omega_{m}$, does not improve even when incorporating the NS EoS. The calculation of luminosity distance is sensitive to $\Omega_{m}$ at high redshifts, while all the GW events involved in this work are distributed in the low-redshift universe. Therefore, the impact of $\Omega_{m}$ on calculating the luminosity distance is negligible for the BNSs detected by the current-generation GW detectors, given their limited horizon distances. As a result, we cannot expect to infer $\Omega_{m}$ from these detections.

Similar to the improved estimation of $H_{0}$ and $m_{\rm max}$, other mass model parameters are also expected to show improvements when considering the tidal information from GW observations due to the additional constraint from $m-\Lambda$ relations. 
The mean of the Gaussian (mass) distribution is inferred more precisely and accurately for $50$ GW events when the NS EoS is considered (see Fig.~\ref{fig:pop_parameters_compare_50events_gaussian}).
However, for $5$ GW events with the same mass distribution, there is no significant impact on the estimation of $\mu$. This is because the sample size of $5$ GW events is too small to reflect the significance of $\mu$ inference from $m-\Lambda$ constraints. 
Similar to $m_{\rm max}$, a strong correlation between $\mu$ and $H_{0}$ is evident for both $5$ and $50$ events~\citep{Taylor:2011fs}. 
While inferring the hyperparameters, as one samples larger $H_{0}$ values, the inferred redshift from the measured luminosity distance of the GW event would also increase, leading to a decrease in the inferred source-frame masses corresponding to an observed detector-frame mass. This implies a negative correlation between estimated $\mu$ and estimated $H_{0}$. 
Consequently, the precision in the measurements of $\mu$ is closely linked to those of $H_{0}$ and $m_{\rm max}$, since the connection among these parameters plays a crucial role for fitting the mass model to the observed data. As a result, $\mu$ is fairly well constrained even when considering the population model alone.
Furthermore, incorporating additional information through $m-\Lambda$ constraints helps improve its precision. This highlights the substantial impact of the NS EoS in diminishing the degree of correlation between $\mu$ and $H_{0}$.
The estimate of the other mass distribution parameter $\sigma$ -- the standard deviation of the Gaussian distribution -- remains similar irrespective of the incorporation of the NS EoS. 
Note that even as many as $50$ events in current-generation detectors may be insufficient in capturing the influence of the EoS on $\sigma$. A thorough investigation of the impact of the NS EoS on $\sigma$ would require either a larger sample of BNS events or next-generation detectors, which is beyond the scope of the present study.

For the double Gaussian mass distribution, the posterior distributions of the parameters $(\mu_{1}, \sigma_{1})$ for the first Gaussian component and $(\mu_{2}, \sigma_{2})$ for the second Gaussian component are expected to exhibit similar characteristics to the posteriors of $(\mu, \sigma)$, observed for the unimodal Gaussian mass distribution. This expectation stems from the nature of the double Gaussian distribution, which is a combination of two distinct Gaussian distributions.
While the posterior of the first peak $\mu_{1}$ of the double Gaussian mass distribution shows no impact from the consideration of the NS EoS (Fig.~\ref{fig:pop_parameters_compare_5events_double_gaussian}) for $5$ GW events, $\mu_{1}$ is constrained with better precision for $50$ GW events when the NS EoS is taken into account. 
Similar to the correlation between $\mu$ and $H_{0}$ for BNS events with the Gaussian mass distribution, there exists a correlation between $\mu_{1}$ and $H_{0}$ for BNS events following the double Gaussian mass distribution for the same reason.
However, due to the presence of fewer GW events near the secondary peak ($\mu_{2}=1.8~M_{\odot}$) of the same mass distribution, even for the population of $50$ events,  there is no significant improvement in the inference of $\mu_{2}$ or its correlation with $H_{0}$.
More events are required to have a significant number of BNSs at the secondary peak to observe a similar qualitative feature as seen for $\mu_{1}$. The other mass parameters, such as $\sigma_{1}$, $\sigma_{2}$, and $w$, do not benefit from the inclusion of the NS EoS parameters. Increasing the number of events allows for more precise constraints of these parameters due to reduced statistical uncertainty. 
Given that only a few BNS events are expected with current-generation detectors~\citep{KAGRA:2021vkt}, these parameters remain poorly constrained in such cases, as already demonstrated in this study using $5$ events.
To further explore the impact of the NS EoS in the inference of these parameters and understand the required number of events, a large number of simulated BNS events needs to be analyzed, which is realistic for the next-generation detector era.
We leave this exercise for future investigations.

\begin{figure*} 
    \centering
    \includegraphics[scale=0.44]{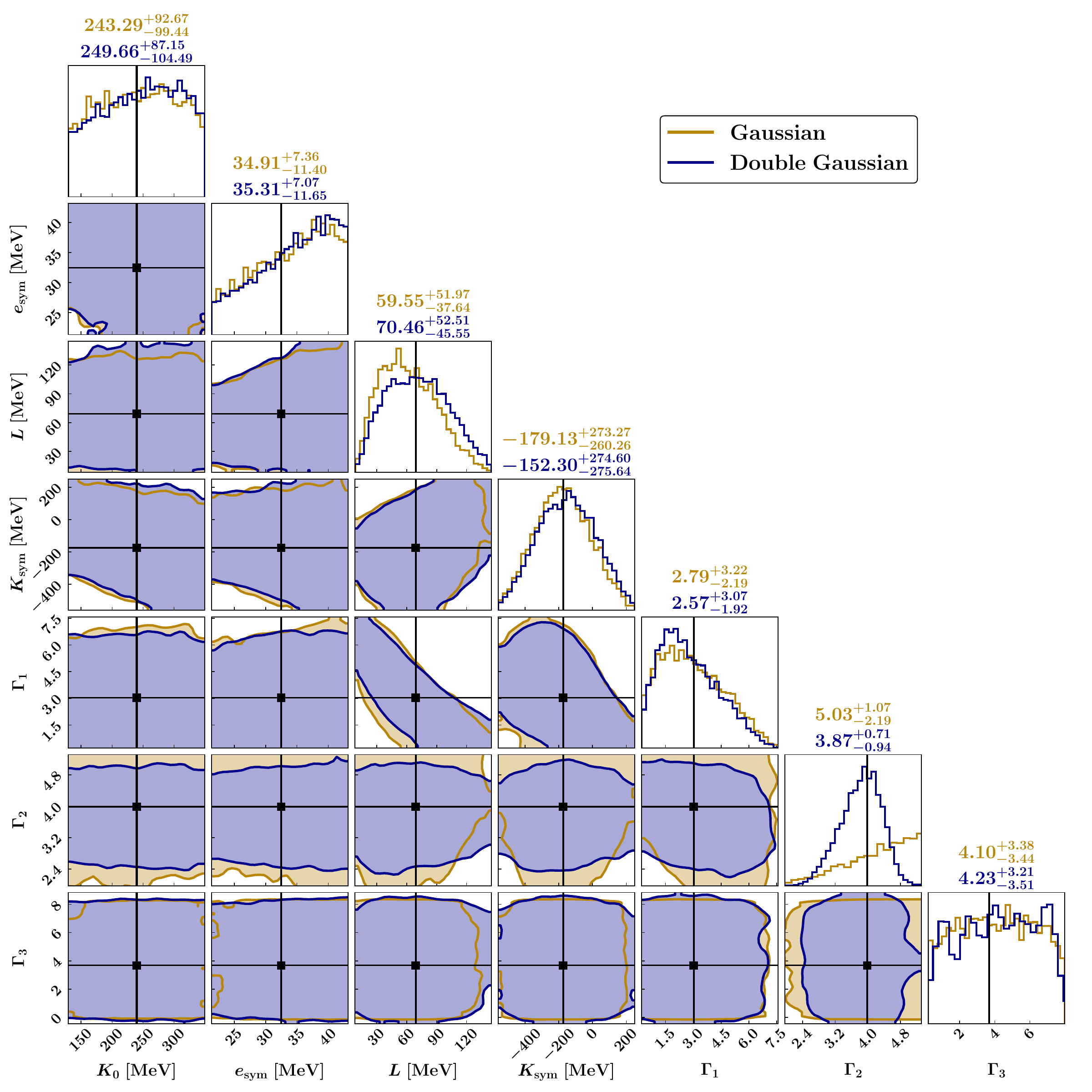}
    \caption{Comparison of the constraints on the NS EoS parameters from $5$ GW events, following the Gaussian and double Gaussian mass distributions. The black solid lines correspond to the true NS EoS parameters. The uncertainty of each parameter corresponding to the $90\%$ credible interval is shown at the top of the $1$D posterior for both mass distributions.}
    \label{fig:EoS_corner_mass_distribution_5}
\end{figure*}

\begin{figure*} 
    \centering
    \includegraphics[scale=0.44]{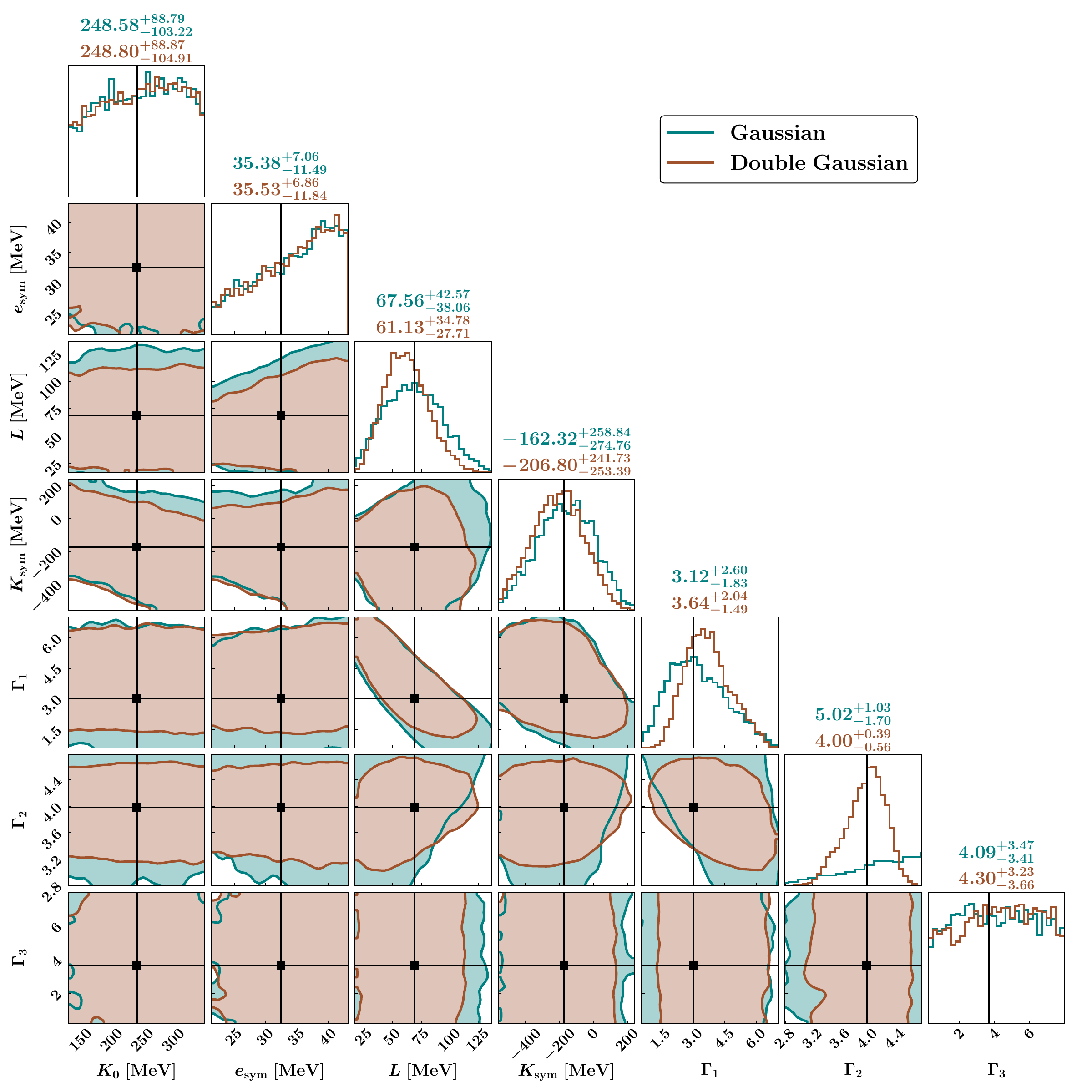}
    \caption{Same as Fig.~\ref{fig:EoS_corner_mass_distribution_5}, but using $50$ events.}
    \label{fig:EoS_corner_mass_distribution_50}
\end{figure*}

The estimation of the redshift evolution parameter is difficult to constrain, irrespective of the choice of the mass distributions and the number of GW events. The redshift evolution parameter $\gamma$ is weakly constrained because the redshift distribution of BNS is primarily limited to smaller luminosity distance $\sim 350$ Mpc. However, the redshift distribution parameter shows an improvement in the precision when the number of detected GW events is increased from $5$ to $50$. Additionally, there is a slight enhancement in inferring $\gamma$ for $50$ GW events following the Gaussian mass distribution when the NS EoS is considered (see Fig.~\ref{fig:pop_parameters_compare_50events_gaussian}). 
Otherwise, we do not observe any significant impact of the NS EoS on inferring the redshift evolution parameter.

\begin{figure*} 
    \centering
    \includegraphics[scale=0.8]{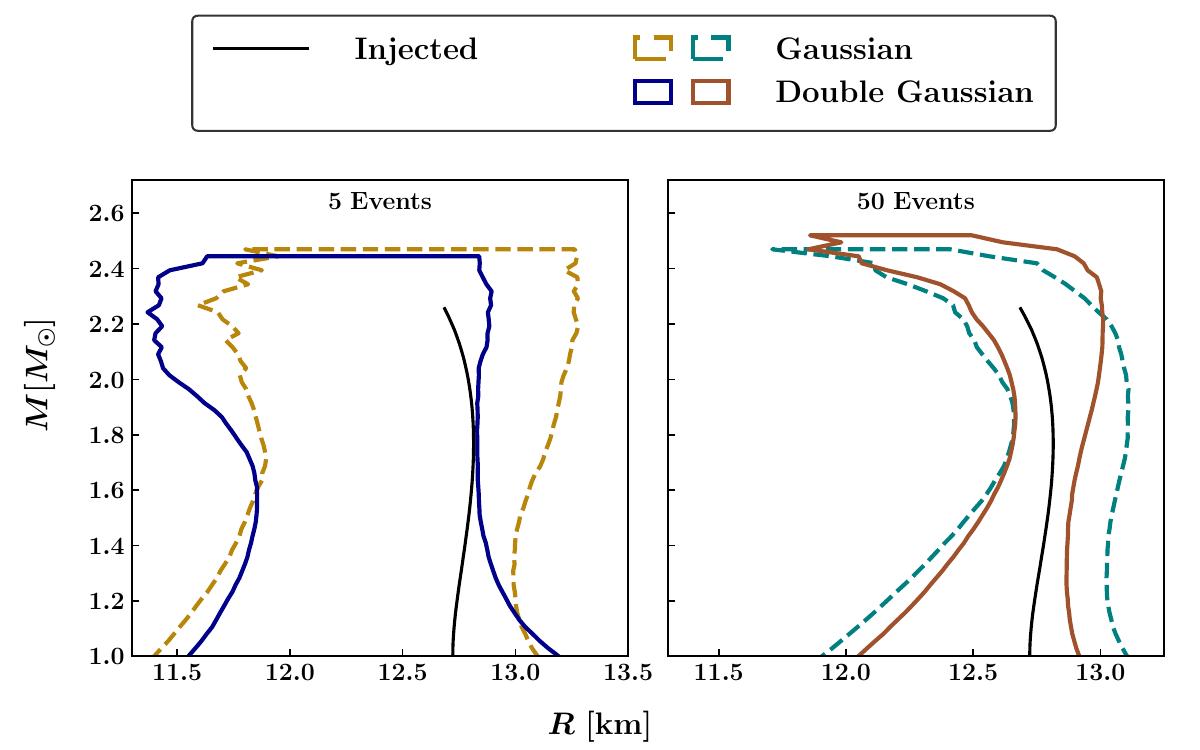}
    \caption{Comparison of the $90\%$ credible regions of mass-radius posteriors between the Gaussian and double Gaussian mass distributions for $5$ (left) and $50$ (right) events. The black solid line indicates the true mass-radius relation corresponding to the injected NS EoS parameters.}
    \label{fig:mr_post_mass_distribution}
\end{figure*}

We find that the NS EoS strongly influences the precision of the inferred
model parameters $H_{0}$ and $m_{\rm max}$, if not $\mu$ (for the Gaussian mass distribution) and $\mu_{1}$ (for the double Gaussian mass distribution).
However, its impact on the other population parameters describing mass and redshift distributions may vary depending on the chosen population model.
These improvements stem from imposing $m-\Lambda$ relations derived from NS EoS into GW observations, which are not considered during the estimation of BNS parameters from GW data. 
Therefore, this study advocates for performing simultaneous inference of population, cosmology, and NS EoS for comprehensive analysis.

\begingroup
\begin{table*}
\begin{center}
\begin{tabular}{| c | c@{\hskip 1.1em} | c@{\hskip 1.1em} | c@{\hskip 1.1em}| c@{\hskip 1.1em} |}
\hline
\multirow{3}{*}{\centering EoS Parameters} & \multicolumn{4}{c|}{Mass Model}\\
\cline{2-5}
& \multicolumn{2}{c|}{Gaussian} & \multicolumn{2}{c|}{Double Gaussian} \\
\cline{2-5}
& $5$ Events & $50$ Events & $5$ Events & $50$ Events  \\
\hline 
$K_{0}$ & $243.29^{+92.67}_{-99.44}$ & $248.58^{+88.79}_{-103.22}$ & $249.66^{+87.15}_{-104.49}$ & $248.80^{+88.87}_{-104.91}$ \\
$e_{\rm sym}$ & $34.91^{+7.36}_{-11.40}$ & $35.38^{+7.06}_{-11.40}$ & $35.31^{+7.07}_{-11.65}$ & $35.53^{+6.86}_{-11.84}$ \\
$L$ & $59.55^{+51.97}_{-37.64}$ & $67.56^{+42.57}_{-38.06}$ & $70.46^{+52.51}_{-45.55}$ & $61.13^{+34.78}_{-27.71}$ \\
$K_{\rm sym}$ & $-179.13^{+273.27}_{-260.26}$ & $-162.32^{+258.84}_{-274.76}$ & $-152.30^{+274.60}_{-275.64}$ & $-206.80^{+241.73}_{-253.39}$ \\
$\Gamma_{1}$ & $2.79^{+3.22}_{-2.19}$ & $3.12^{+2.60}_{-1.83}$ & $2.57^{+3.07}_{-1.92}$ & $3.64^{+2.04}_{-1.49}$ \\
$\Gamma_{2}$ & $5.03^{+1.07}_{-2.19}$ & $5.02^{+1.03}_{-1.70}$ & $3.87^{+0.71}_{-0.94}$ & $4.00^{+0.39}_{-0.56}$ \\
$\Gamma_{3}$ & $4.10^{+3.38}_{-3.44}$ & $4.09^{+3.47}_{-3.41}$ & $4.23^{+3.21}_{-3.51}$ & $4.30^{+3.23}_{-3.66}$ \\
\hline
\end{tabular}
\caption{\label{tab:eos_posterior} $90\%$ credible intervals representing the uncertainty in measuring the EoS parameters from GW events of $5$ and $50$, following the Gaussian and double Gaussian mass distributions.}
\end{center}
\end{table*}
\endgroup

\subsection{Constraint of NS EoS Parameters} \label{subsec:result_2}

The inferred NS EoS parameters from $5$ ($50$) events, following the Gaussian and double Gaussian mass distributions, are shown in Fig.~\ref{fig:EoS_corner_mass_distribution_5} (Fig.~\ref{fig:EoS_corner_mass_distribution_50}).
The corresponding $90\%$ intervals are listed in Table~\ref{tab:eos_posterior}. It should be emphasized that we infer the NS EoS parameters together with the population and the cosmological parameters.~\footnote{To reduce clutter in the figures, we show the full joint posterior distributions of all the hyperparameters in  Fig.~\ref{fig:joint_post_gaussian} of Appendix~\ref{appendix:joint_post} -- for $50$ events following the Gaussian mass distribution.}
We do not observe any significant correlations between the NS EoS parameters, on the one hand, and the population (except for $m_{\rm max}$) or the Hubble constant, on the other.

Among the nuclear EoS parameters, $K_{0}$ and $e_{\rm sym}$ do not show any improvement with the increased number of GW events, regardless of the mass distribution (see Figs.~\ref{fig:EoS_corner_gaussian} and~\ref{fig:EoS_corner_double_gaussian} in Appendix~\ref{appendix:eos_compare_5_50} for reference). The posteriors of these parameters are dominated by the effective priors of the EoS parameters, as shown in Fig.~\ref{fig:EoS_prior_corner} of Appendix~\ref{appendix:EoS_effective_prior}. 
The posterior distributions of $K_{\rm sym}$ exhibit a slight improvement compared to the corresponding effective prior (see Fig.~\ref{fig:EoS_prior_corner}).
The only nuclear parameter to be well constrained by these simulated GW observations is $L$.
On the other hand, PP parameters $\Gamma_{1}$ and $\Gamma_{2}$ are well constrained, but $\Gamma_{3}$ is not constrained. 
This holds true irrespective of the number of GW events and the mass distribution of NSs. It has been demonstrated in Ref.~\cite{Biswas:2024hja} that most EoSs reach their maximum masses at densities lower than those where the last polytropic indices are attached. Hence, we cannot constrain $\Gamma_3$. In other words, the hybrid nuclear+PP EoS parameterization can be modeled with fewer polytropic indices, thereby reducing computational time without significantly affecting the results.

Since the maximum mass of a NS is governed by its EoS, constraining the NS EoS parameters inherently provides the estimation of $m_{\rm max}$. In our study, we observe that the well-constrained EoS parameters $(L, \Gamma_{1}, \Gamma_{2})$ exhibit correlations with $m_{\rm max}$, allowing for its tighter bounding than the population-only analysis, as discussed in Sec.~\ref{subsec:result_1}.
For instance, with only a few GW events ($5$ events), $m_{\rm max}$ cannot be inferred accurately since there is no simulated BNS close enough to the maximum mass, in either mass distribution. Even with $50$ BNSs following the Gaussian mass distribution, $m_{\rm max}$ remains weakly constrained due to the absence of high-mass mergers--none of the source-frame NS masses exceed $1.51~M_{\odot}$ (see the left panel of the top row of Fig.~\ref{fig:mass_hist}). 
Even so, our method improves the constraints on $m_{\rm max}$ compared to the spectral siren method.  
Nonetheless, NS EoS into information improves the constraint on $m_{\rm max}$ by a factor of $\gtrsim 2$. 

In contrast, the double Gaussian mass distribution naturally allows a few NSs to occur near the maximum mass region
(see Fig.~\ref{fig:mass_hist}), which leads to a better estimate of $m_{\rm max}$ compared to the Gaussian mass distribution, irrespective of whether the NS EoS is employed or not in the analysis. 
Though it suggests that this improvement is due to the intrinsic nature of mass models, the current sample size of $50$ events may not be sufficient to draw definitive conclusions. To robustly assess the impact of the NS EoS on constraining $m_{\rm max}$, a larger number of events would be needed. However, if the current samples are assumed to be representative of their respective mass distributions, the observed $\sim 3$–$4$ times improvement in constraining $m_{\rm max}$ for the double Gaussian model, compared to the Gaussian model, can be attributed to the intrinsic characteristics of the underlying population.

We present the $90\%$ credible regions of the inferred mass-radius distributions in Fig.~\ref{fig:mr_post_mass_distribution}, which shows the impact of the mass distribution on inferring the NS EoS parameters, both for $5$ and $50$ GW events. 
The posteriors of the NS EoS parameters from $5$ ($50$) events, following the Gaussian and double Gaussian mass distributions, are shown in Fig.~\ref{fig:EoS_corner_mass_distribution_5} (Fig.~\ref{fig:EoS_corner_mass_distribution_50}). 
The comparison of mass-radius plots, as shown in Fig.~\ref{fig:mr_post_mass_distribution}, between the Gaussian and double Gaussian mass distributions for BNS events, reveals that the mass-radius is not well constrained near the maximum mass for BNSs following the Gaussian mass distribution compared to the double Gaussian mass distribution. This result holds for both $5$ and $50$ events.
BNS events following the double Gaussian mass distribution can relatively well constrain the mass-radius region, as shown in Fig.~\ref{fig:mr_post_mass_distribution}, irrespective of the number of events. This is expected because the NS EoS parameters $\Gamma_{1}$ and $\Gamma_{2}$ are constrained with better precision from the BNS events with the double Gaussian mass distribution compared to the Gaussian mass distribution.
In particular, the presence of higher mass NSs in the double Gaussian mass distribution, compared to the Gaussian mass distribution (see, Fig.~\ref{fig:mass_hist}), contributes to better constraining some NS EoS parameters and hence the mass-radius of NSs. 
This suggests that the nature of the true astrophysical mass distribution influences the precision with which the model parameters are constrained.

\section{Conclusion} \label{sec:conclusion}

In this work, we demonstrated the efficacy of the simultaneous inference of population, cosmology, and NS EoS using simulated BNS mergers observed in GWs.
This approach applies the spectral siren method exclusively to a population of BNS signals by comparing the source-frame mass spectrum with the observed mass distribution.
While doing so, it also utilizes the measured NS tidal parameters along with the $m-\Lambda$ constraints to estimate the population and cosmological parameters.
Our study shows that the spectral siren method can constrain the Hubble constant to within $\Delta H_{0}\sim 70-80$~\kmsMpc (see Table~\ref{tab:pop_posterior}) when analyzing $50$ events distributed over $\sim 350$ Mpc.
Moreover, $m_{\rm max}$ remains poorly constrained due to the paucity or 
absence of BNS events near the maximum NS mass region of our simulations.
Incorporating the NS EoS parameters into the spectral siren method explicitly imposes the $m-\Lambda$ constraints through the NS EoS by including the tidal parameters of BNSs with their observed masses and luminosity distances. This additional $m-\Lambda$ relation, in conjunction with cosmology (especially $H_{0}$), helps to break the mass-redshift degeneracy in the observed BNS parameters of GW data.
Thus including tidal parameters in the spectral siren method significantly improves the inference of key parameters, such as, $H_{0}$ and $m_{\rm max}$, as well as certain population parameters, such as $\mu$ and $\mu_{1}$ corresponding to the Gaussian and double Gaussian mass distributions, respectively.
These results highlight the importance of inferring $H_{0}$ alongside population and NS EoS parameters. 
In contrast, improvements in the measurement of other parameters, such as ($\sigma$, $\gamma$, $\Omega_{m}$) for Gaussian and  ($\sigma_{1}$, $\mu_{2}$,$\sigma_{2}$,$\gamma$, $\Omega_{m}$) for double Gaussian mass distribution, are not comparatively significant.
This is due to significant uncertainties in the BNS parameters and the limited number of BNS mergers -- as expected for the current-generation detectors.
In future-generation detectors, where the precise measurements of BNS parameters for a large number of events will cause systematics to dominate over statistical uncertainties, this method will become more useful.
Neglecting the $m-\Lambda$ relations from the NS EoS in such cases can bias the inferred model parameters. However, this bias will be subdominant to the statistical errors for current-generation detectors, given the limited number of BNS events expected to be observed.
Nonetheless, the method is essential for estimating the Hubble constant from BNS events observed as dark sirens by current-generation detectors.

This study of simulated BNS events shows that most nuclear parameters in the EoS model are not constrained by the GW events observed with the current-generation detectors. However, one can employ informed priors over $K_{0}$, $e_{\rm sym}$, and $L$, derived from laboratory-based nuclear experiments, e.g., PREX~\citep{PREX:2021umo}, CREX~\citep{CREX:2022kgg}, and theoretical predictions performed by chiral effective field theory~\citep{Drischler:2020yad}, to improve the precision of estimating model parameters.
It is also of interest for future studies to investigate how different (nuclear) EoS models influence improvements in the inference of hyperparameters describing the population, cosmology, and the NS EoS.
This approach also updates the constraints on nuclear parameters obtained thus far from GW observations. Moreover, an informed prior by combining other observations related to the NS properties, such as mass-radius measurements~\citep{Riley:2019yda, Miller:2019cac, Riley:2021pdl, Miller:2021qha} from X-ray observations and maximum mass thresholds from radio pulsar measurements~\citep{Fonseca:2021wxt}, can aid in precisely measuring the Hubble constant. Thus, our method may enable accurate and precise determination of the Hubble constant by integrating data from nuclear experiments and various astrophysical observations, including X-ray and radio data, with GW data.

The method proposed in this study can be generalized to any parameterized model, encompassing population, cosmology, and NS EoS.
We consider that both components of BNS follow the same mass distribution. However, there is also evidence from galactic BNS observations that they may come from separate mass distributions~\citep{Farrow:2019xnc}. 
It is also important to consider realistic spin models, which are not considered in this study. Spin mismodeling can lead to biased estimation of mass distributions~\citep{Biscoveanu:2021eht}.
Our method can easily be extended to incorporate different parameterized population models in the Bayesian formalism. The precision and accuracy of inferring different hyperparameters may vary depending on the choice of models.
We have also made a simplified assumption regarding the redshift evolution model as a power law with $\gamma=0$ as an injected parameter~\citep{KAGRA:2013rdx, LIGOScientific:2020kqk}. Even for real observations with the current-generation detectors, the power-law distribution is reasonable, as all the BNSs are distributed within the low-redshift Universe. The choice of $\gamma$ may impact the accuracy and precision of estimated model parameters. However, it is difficult to infer the redshift evolution parameter with significant precision and accuracy with a small number of BNS events, which is expected with current-generation detectors due to the smaller horizon redshift.

In this work, we consider a network of $3$ detectors, consisting of two LIGO detectors and the Virgo detector.
With the addition of KAGRA~\citep{KAGRA:2020agh} and LIGO-India~\citep{Saleem:2021iwi} to the existing detector network, the detection rate of BNS mergers is expected to increase, accompanied by improved precision in the inferred BNS parameters and an extended horizon redshift. These are achievable due to the overall improvement in network SNR and network duty factor, where by network we mean one involving at least 3 detectors, which is the type of multi-detector study we conducted here. In particular, a larger detector network helps reduce the degeneracy between the luminosity distance and the inclination angle~\citep{Usman:2018imj}. 
Consequently, the aforementioned detector network is likely to make more precise measurements of key parameters such as component masses, tidal deformabilities, and luminosity distance, which can result in tighter constraints on the Hubble constant. Furthermore, the increased horizon distance may improve the estimation of redshift evolution parameter(s).

Projecting further regarding next-generation ground-based detectors, such as the CE~\citep{Evans:2021gyd} and the ET~\citep{Maggiore:2019uih}, which are planned for the next decade, a large number of BNSs $\sim 10^{5}$ are expected to be detected per year~\citep{Branchesi:2023mws}. 
Since only a small fraction of the total events is expected to be bright standard sirens up to a relatively small redshift of $z\sim 2$~\citep{Mpetha:2022xqo}, compared to the horizon redshift of $z\sim 10$ for the next-generation detectors, most of the detected BNS mergers will remain dark sirens. This is due to the coverage limitation of the future spectroscopic galaxy surveys~\citep{2011arXiv1110.3193L}. 
In such circumstances, the method discussed in this paper is particularly beneficial for probing cosmology as well as mass and redshift distributions. Since the horizon redshift is quite large, this method also enables the probing of different cosmological parameters, such as the matter density and the dark energy equation of state, which are not possible with the current-generation detectors. 
One must be careful to account for the impact of lensing~\citep{Canevarolo:2024muf} of GW signals while performing our method for the next-generation detectors to ensure the unbiased estimation of NS EoS, population, and cosmology. However, this concern is not important for the current-generation detectors, due to their small horizon distance of $d_{L}\sim 350$~Mpc.  However, it is not trivial to estimate population, NS EoS, and cosmology from the observations of dark BNS mergers, some of which may be lensed. The incorporation of EoS may mitigate the systematics due to lensing in inferring different model parameters, which can be explored in future studies.

\acknowledgments

The authors would like to thank Simone Mastrogiovanni for carefully reading the manuscript and sharing his comments. T.G. also thanks Surhud More for some useful discussion on this study.
T.G. acknowledges support from JSPS Grant-in-Aid for Transformative Research Areas (A) No. 23H04893.
T.G. gratefully acknowledges the computational facilities provided by IUCAA, including the LDG cluster Sarathi and Pegasus, as well as the use of the LDG cluster Hawk at Cardiff University, supported by STFC grants ST/I006285/1 and ST/V005618/1.
B.B.  acknowledges support from the Knut and Alice Wallenberg Foundation under grant Dnr.~KAW~2019.0112 and the Deutsche Forschungsgemeinschaft (DFG, German Research Foundation) under Germany's Excellence Strategy-EXC~2121 ``Quantum Universe'' –
390833306. S. B. acknowledges support from the NSF under grant PHY-2309352. S.J.K gratefully acknowledges support from SERB grants SRG/2023/000419 and MTR/2023/000086.

\appendix

\section{Bayesian Framework} \label{appendix_bayesian_formalism}

In this appendix, we briefly review the hierarchical Bayesian framework used to infer model hyperparameters $\bm{\Theta}$ from GW data. The core concept involves modeling the population~\footnote{This approach also entails estimating NS EoS and cosmological parameters while modeling the population from BNS observations.}
assuming a conditional prior $p(\theta \mid \bm{\Theta} )$. This prior characterizes, e.g., the mass-redshift distribution of binary merger parameters $\bm{\theta}$, given the hyperparameters $\bm{\Theta}$, which determine the shape of the prior distribution.

According to Bayes's theorem, we can write the posterior probability distribution of $\bm{\Theta}$ for a set of GW data $\{d\}$ as follows:

\begin{equation}
    p(\bm{\Theta} \mid \{d\}) = \frac{\mathcal{L}(\{d\} \mid \bm{\Theta})p(\bm{\Theta})}{Z(\{d\})} \,,
\end{equation}
where $\mathcal{L}(\{d\} \mid \bm{\Theta})$ denotes the joint likelihood, which is just the product of the individual event likelihoods,
\begin{equation}
    \mathcal{L}(\{d\} \mid \bm{\Theta})= \prod_{i} \mathcal{L}(d_{i} \mid \bm{\Theta}) \,,
\end{equation}
assuming that the data are statistically independent.
Here, $i$ denotes the $i$th GW event. For the rest of the calculation, we drop the use of $i$ for notational convenience.
The individual event likelihood $\mathcal{L}(d \mid \bm{\Theta})$ is computed from measurements of the BNS parameters $\bm{\theta}$: 

\begin{equation}
    \mathcal{L} (d \mid \bm{\Theta}) = \int \mathcal{L} (d \mid \bm{\theta}) p(\bm{\theta} \mid \bm{\Theta}) d\bm{\theta}
\end{equation}
Here, the second term $p(\bm{\theta} \mid \Theta)$ represents the model prior. This equation can be further simplified by considering different model parameters $\bm{\Theta}=\{\bm{\Theta}_{\mathcal{E}}, \bm{\Theta}_{m}, \bm{\Theta}_{z}, \bm{\Theta}_{c}\}$ and BNS parameters $\bm{\theta} = \{m_{1}, m_{2}, z, \Lambda_{1}, \Lambda_{2}\}$ explicitly,

\begin{widetext}
\begin{equation} \label{eq:likelihood_explicit}
    \mathcal{L} (d \mid \bm{\Theta}) = p(\bm{\Theta}) \int dm_{1} \int dm_{2} \int dz \int d\Lambda_{1} d\Lambda_{2} ~\mathcal{L} (d \mid \bm{\theta}) p(m_{1}, m_{2}, \Lambda_{1}, \Lambda_{2} \mid \bm{\Theta}_{m}, \bm{\Theta}_{\mathcal{E}}) p(z \mid \bm{\Theta}_{z})  \,,
\end{equation}
\end{widetext}
where $p(\bm{\Theta}) =  p(\bm{\Theta_{m}}) p(\bm{\Theta_{z}}) p(\bm{\Theta_{c}}) p(\bm{\Theta_{\mathcal{E}}})$ denotes the prior over model hyperparameters $\bm{\Theta}$. Given that the tidal deformability is mass dependent for a specific equation of state $\bm{\Theta}_{\mathcal{E}}$, Eq.~\eqref{eq:likelihood_explicit} can be further simplified when considering NS tidal parameters,

\begin{widetext}
\begin{equation}
    \mathcal{L} (d \mid \bm{\Theta}) = p(\bm{\Theta}) \int dm_{1} \int dm_{2} \int dz ~\mathcal{L} (d \mid m_{1}, m_{2}, z, \Lambda_{1}(m_{1}, \bm{\Theta}_{\mathcal{E}}), \Lambda_{2}(m_{2}, \bm{\Theta}_{\mathcal{E}}) ) p(m_{1}, m_{2}\mid \bm{\Theta}_{m}) p(z \mid \bm{\Theta}_{z})  \,,
\end{equation}
\end{widetext}

since one has

\begin{align} 
     p(m_{1}, & m_{2},  \Lambda_{1}, \Lambda_{2} \mid \bm{\Theta}_{m}, \bm{\Theta}_{\mathcal{E}}) = p(m_{1}, m_{2} \mid \bm{\Theta}_{m}) \nonumber \\
     &\times\delta(\Lambda_{1}-\Lambda_{1}(m, \bm{\Theta}_{\mathcal{E}})) \delta(\Lambda_{2}-\Lambda_{2}(m, \bm{\Theta}_{\mathcal{E}})) \,.
\end{align}

It is important to note that we do not incorporate $\bm{\Theta_{\mathcal{E}}}$ when we only focus on population and cosmology, as already mentioned in Sec.~\ref{sec:method}. In that case, we exclude tidal parameters from the set of BNS parameters.

\section{Jacobian Calculation for Frame Transformation} \label{appendix:jacobian} 

The Jacobian corresponding to the coordinate transformation from $(\mathcal{M}_{c}^{z}, q, d_{L})$ to $(m_{1}, m_{2}, z)$, as shown in Eq.~\eqref{eq:jacobian_matrix}, can be simplified as follows:

\begin{eqnarray} \label{eq:jacobian_cal}
    J \left( \frac{\mathcal{M}_{c}^{z}, q, d_{L}}{m_{1}, m_{2}, z}\right) && = 
    \begin{bmatrix}
  \frac{\partial \mathcal{M}_{c}^{z}}{\partial m_{1}} & 
    \frac{\partial \mathcal{M}_{c}^{z}}{\partial m_{2}} & 
    \frac{\partial \mathcal{M}_{c}^{z}}{\partial z} \\[2.5ex] 
  \frac{\partial q}{\partial m_{1}} & 
    \frac{\partial q}{\partial m_{2}} & 
    \frac{\partial q}{\partial z} \\[2.5ex]
  \frac{\partial d_{L}}{\partial m_{1}} & 
    \frac{\partial d_{L}}{\partial m_{2}} & 
    \frac{\partial d_{L}}{\partial z} 
    \end{bmatrix} \nonumber \\
    && =
    \begin{bmatrix}
  \frac{\partial \mathcal{M}_{c}^{z}}{\partial m_{1}} & 
    \frac{\partial \mathcal{M}_{c}^{z}}{\partial m_{2}} & 
    \frac{\partial \mathcal{M}_{c}^{z}}{\partial z} \\[2.5ex] 
  \frac{\partial q}{\partial m_{1}} & 
    \frac{\partial q}{\partial m_{2}} & 
    0 \\[2.5ex]
  0 & 
    0 & 
    \frac{\partial d_{L}}{\partial z} 
    \end{bmatrix} \nonumber \\
    && = (1+z) \frac{\partial d_{L}}{\partial z} \nonumber  \\
    && \times \left(\frac{\partial \mathcal{M}_{c}}{\partial m_{1}} \frac{\partial q}{\partial m_{2}} - \frac{\partial \mathcal{M}_{c}}{\partial m_{2}} \frac{\partial q}{\partial m_{1}} \right)
\,.
\end{eqnarray}
Evaluating each term in Eq.~\eqref{eq:jacobian_cal}, 
$$\frac{\partial \mathcal{M}_{c}}{\partial m_{1}} = \frac{3\mathcal{M}_{c}}{5m_{1}} - \frac{\mathcal{M}_{c}}{5(m_{1}+m_{2})}~,$$
$$ \frac{\partial q}{\partial m_{1}} = -\frac{m_{2}}{m_{1}^{2}}~,$$
$$\frac{\partial \mathcal{M}_{c}}{\partial m_{2}} = \frac{3\mathcal{M}_{c}}{5m_{2}} - \frac{\mathcal{M}_{c}}{5(m_{1}+m_{2})}~,$$
$$ \frac{\partial q}{\partial m_{2}} = \frac{1}{m_{1}}~,$$
$$\frac{\partial d_{L}}{\partial z} = d_{c}+\frac{c(1+z)}{H(z)}~,$$
and simplifying yields the following final expressions:
\begin{eqnarray}
    \left\lvert J \left( \frac{\mathcal{M}_{c}^{z}, q, d_{L}}{m_{1}, m_{2}, z}\right) \right\rvert && = (1+z)\left[d_{c}+\frac{c(1+z)}{H(z)}\right]\frac{\mathcal{M}_{c}}{m_{1}^{2}} \nonumber \\
    && = (1+z)\left[d_{c}+\frac{c(1+z)}{H(z)}\right] \nonumber \\
    && \times \left[\frac{(m_{1} m_{2})^{3/5}}{m_{1}^{2} (m_{1}+m_{2})^{1/5}}\right] \,.
\end{eqnarray}

\section{Posteriors of Single GW Event} \label{appendix:m_lambda_posterior}

\begin{figure*}
    \centering
    \includegraphics[scale=0.58]{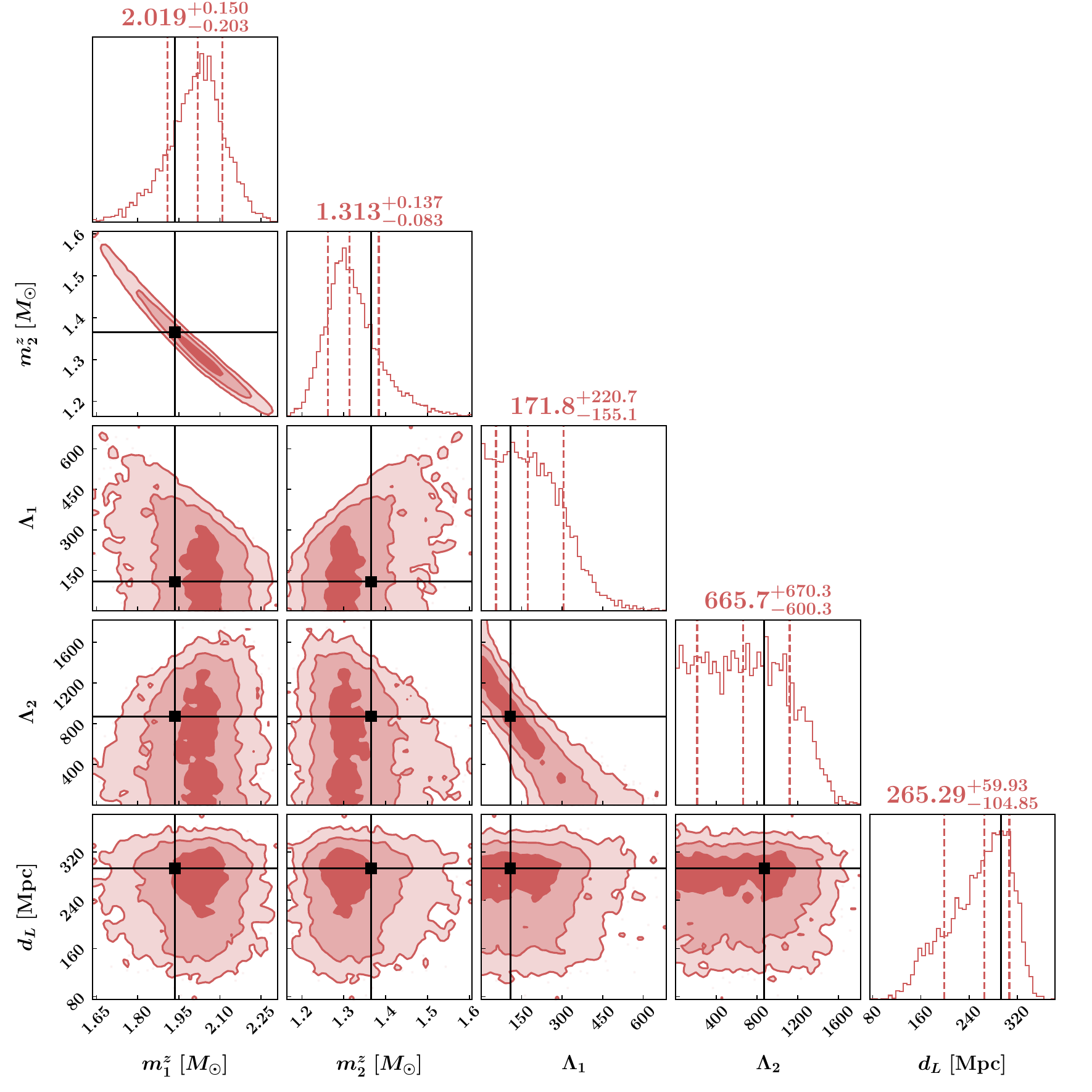}
    \caption{Marginalized joint posteriors of redshifted component masses ($m_{1}^{z}$ and $m_{2}^{z}$), their corresponding tidal deformabilities ($\Lambda_{1}$ and $\Lambda_{2}$), and luminosity distance ($d_{L}$) inferred from GW data of a BNS. The black vertical and horizontal solid lines represent the injected values of the corresponding parameters.}
    \label{fig:m_lambda_corner}
\end{figure*}

In this appendix, we show the posterior distributions of the BNS parameters of a BNS merger in Fig.~\ref{fig:m_lambda_corner}. Though we perform sampling in the observed chirp mass and mass ratio to estimate the source parameters for computational efficiency, we show the component masses (in the detector frame) to illustrate that the posteriors of the BNS parameters exhibit no correlation between mass and tidal parameters, as determined by the NS EoS.

\section{Calculation of Neutron Star Radius and Tidal Deformability} \label{appendix:tov}

The structure of NS is described by the TOV equations~\citep{PhysRev.55.364, PhysRev.55.374}, which are derived from the general relativistic equations for hydrostatic equilibrium. With $G = 1$ and $c = 1$, the TOV equations simplify to
\begin{equation}
\frac{dp(r)}{dr} = -\frac{(\epsilon(r) + p(r))(m(r) + 4 \pi r^3 p(r))}{r (r - 2 m(r))}
\end{equation}
\begin{equation}
\frac{dm(r)}{dr} = 4 \pi r^2 \epsilon(r)
\end{equation}
where $p(r)$ is the pressure, $\epsilon(r)$ is the energy density, and $m(r)$ is the mass enclosed within the radius $r$. These equations are solved with an appropriate equation of state (EoS) that relates the pressure to the energy density.
The radius $R$ of the NS is determined by the point at which the pressure drops to zero, i.e., $p(R) = 0$.

The tidal deformability~\citep{Hinderer:2007mb, Binnington:2009bb, Damour:2009vw} $\Lambda$ of an NS is a measure of its deformation in response to an external tidal field and is defined as
\begin{equation}
\Lambda = \frac{2}{3} k_2 \left( \frac{R}{M} \right)^5\,,
\end{equation}
where $k_2$ is the second Love number, $R$ is the radius, and $M$ is the mass of the NS. The second Love number $k_2$ is calculated using
\begin{equation}
\begin{aligned}
k_2 = \frac{8 C^5}{5} (1 - 2C)^2 \left[ 2 + 2C(y_R - 1) - y_R \right] \\
\times \left\{ 2C \left[ 6 - 3y_R + 3C(5y_R - 8) \right] \right. \\
\left. + 4C^3 \left[ 13 - 11y_R + C(3y_R - 2) + 2C^2(1 + y_R) \right] \right. \\
\left. + 3(1 - 2C)^2 \left[ 2 - y_R + 2C(y_R - 1) \right] \log(1 - 2C) \right\}^{-1}\,,
\end{aligned}
\end{equation}
where $C = \frac{M}{R}$ is the compactness parameter and $y_R$ is determined by solving the following differential equation simultaneously with the TOV equations,
\begin{equation}
\begin{aligned}
\frac{dy(r)}{dr} = -\frac{y(r)^2}{r} - \frac{y(r)}{r} \left[ 1 + 4 \pi r^2 \left( p(r) - \epsilon(r) \right) \right] \\
- \frac{4 \pi r^2 \left( 5 \epsilon(r) + 9 p(r) + \frac{\epsilon(r) + p(r)}{\partial p(r) / \partial \epsilon(r)} - \frac{6}{4 \pi r^2} \right)}{r - 2m(r)} \\
+ \frac{4 (m(r) + 4 \pi r^3 p(r))}{r (r - 2m(r))}
\end{aligned}
\end{equation}
with the boundary condition $y(0) = 2$. The value of $y_R$ is obtained at the star's surface, $r = R$.

By solving the TOV equations along with the above differential equation for $y(r)$, one can obtain the radius $R$ and the Love number $k_2$. Subsequently, these quantities can be used to calculate the tidal deformability $\Lambda$.

\section{Effective Priors over NS EoS Parameters} \label{appendix:EoS_effective_prior}

\begin{figure*} 
    \centering
    \includegraphics[scale=0.39]{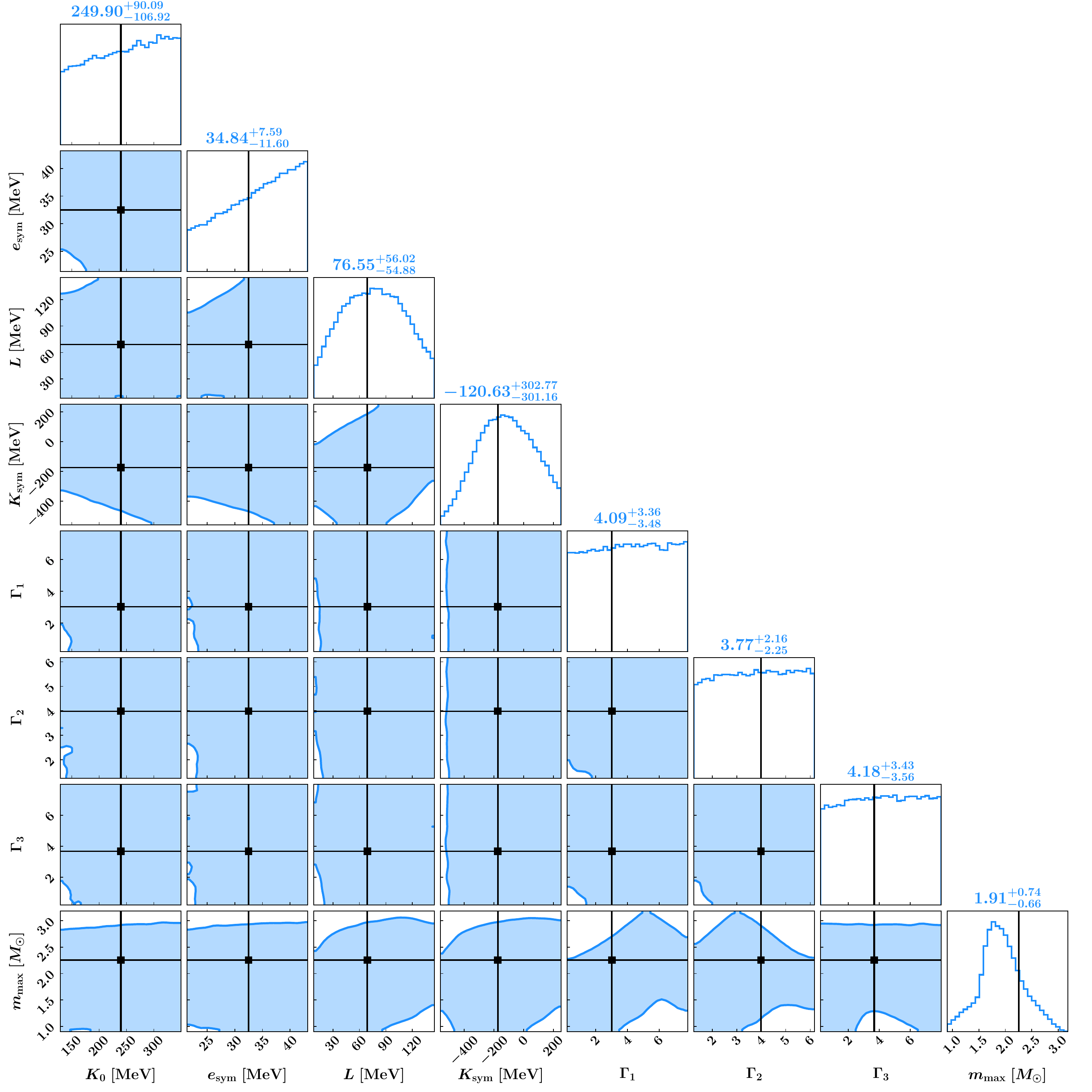}
    \caption{Effective priors for the EoS parameters after imposing the causality constraint. The black solid lines represent the injected NS EoS parameters.}
    \label{fig:EoS_prior_corner}
\end{figure*}

We employ uniform priors for the EoS parameters, as detailed in Table~\ref{tab:mass_model_parameters}. However, not all parameter combinations are physically viable, as a crust-core junction density must be identified. As a result, some parameter combinations are redundant. Given that the fixed BPS crust is matched with the empirical parameterization, the effective priors for the EoS parameters, particularly the empirical parameters, are no longer uniformly distributed, as shown in Fig.~\ref{fig:EoS_prior_corner}.

\section{Joint Posteriors of all Hyperparameters} \label{appendix:joint_post}

\begin{figure*}
    \centering
    \includegraphics[scale=0.24]{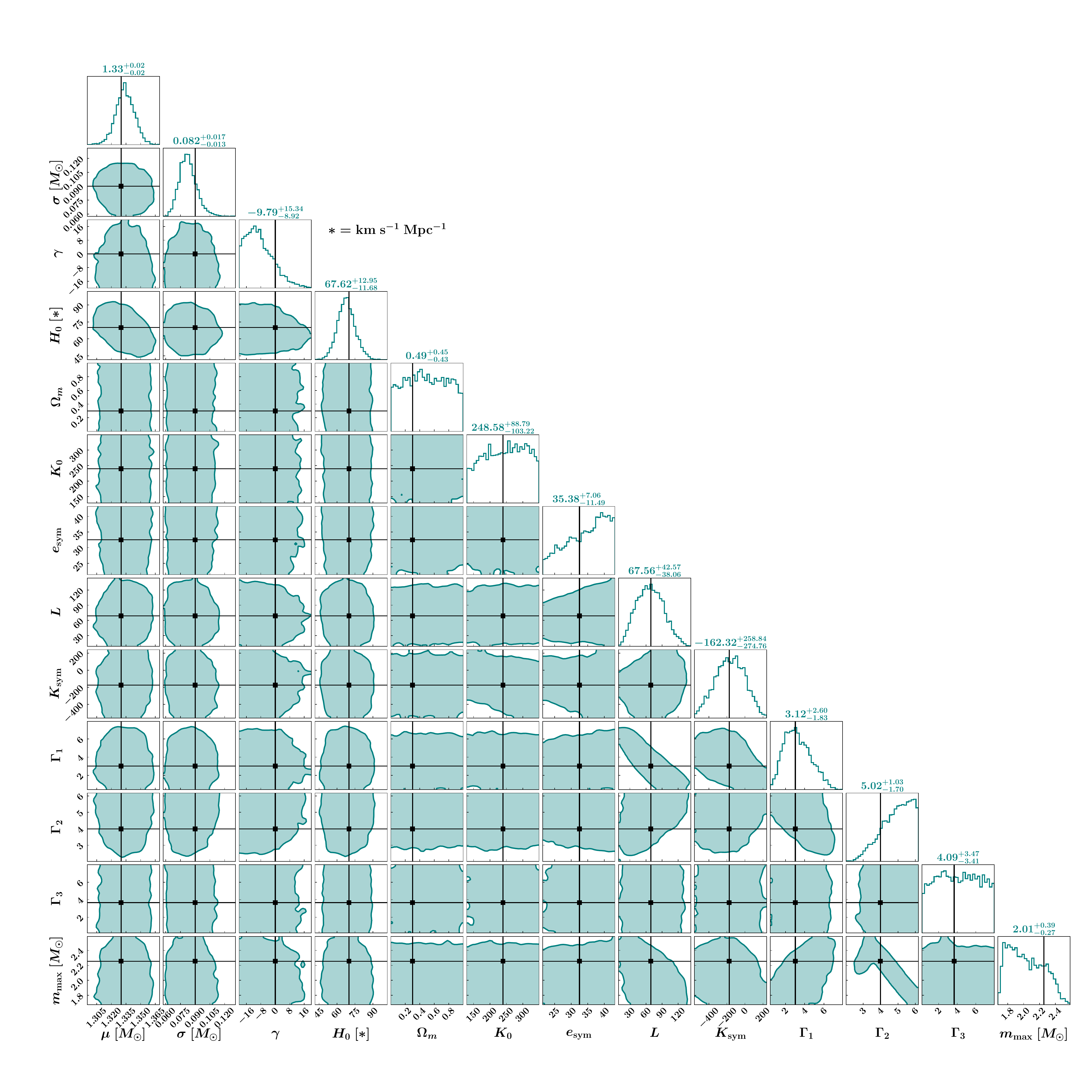}
    \caption{Joint posteriors of all hyperparameters describing the population, cosmology, and NS EoS, inferred from 50 events following the Gaussian mass distribution. The black solid lines indicate the true injected values. The $90\%$ credible interval for each parameter is shown above the corresponding 1D posterior.}
    \label{fig:joint_post_gaussian}
\end{figure*}

In Fig.~\ref{fig:joint_post_gaussian}, we present the joint posterior distributions of hyperparameters describing the population, cosmology, and NS EoS, inferred from $50$ simulated events following the Gaussian mass distribution.

\section{Comparison of NS EoS Parameters Constraints from $5$ and $50$ Events} \label{appendix:eos_compare_5_50}

\begin{figure*} 
    \centering
    \includegraphics[scale=0.44]{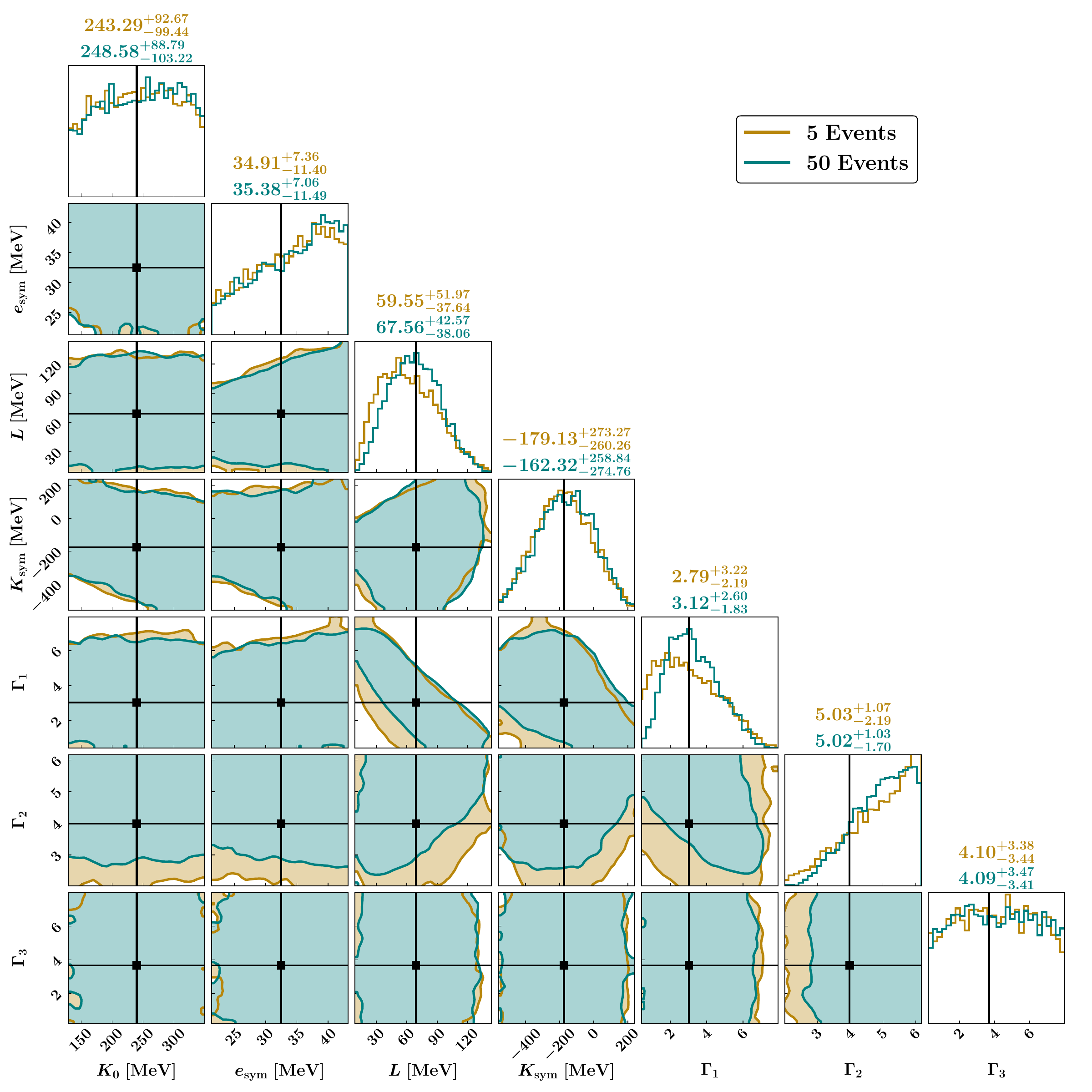}
    \caption{Comparison of the constraints on the NS EoS parameters between $5$ and $50$ GW events, following the Gaussian mass distribution. The black solid lines correspond to the true NS EoS parameters. The uncertainty of each parameter, corresponding to the $90\%$ credible interval, is shown at the top of the respective marginalized $1$D posterior.}
    \label{fig:EoS_corner_gaussian}
\end{figure*}

\begin{figure*} 
    \centering
    \includegraphics[scale=0.44]{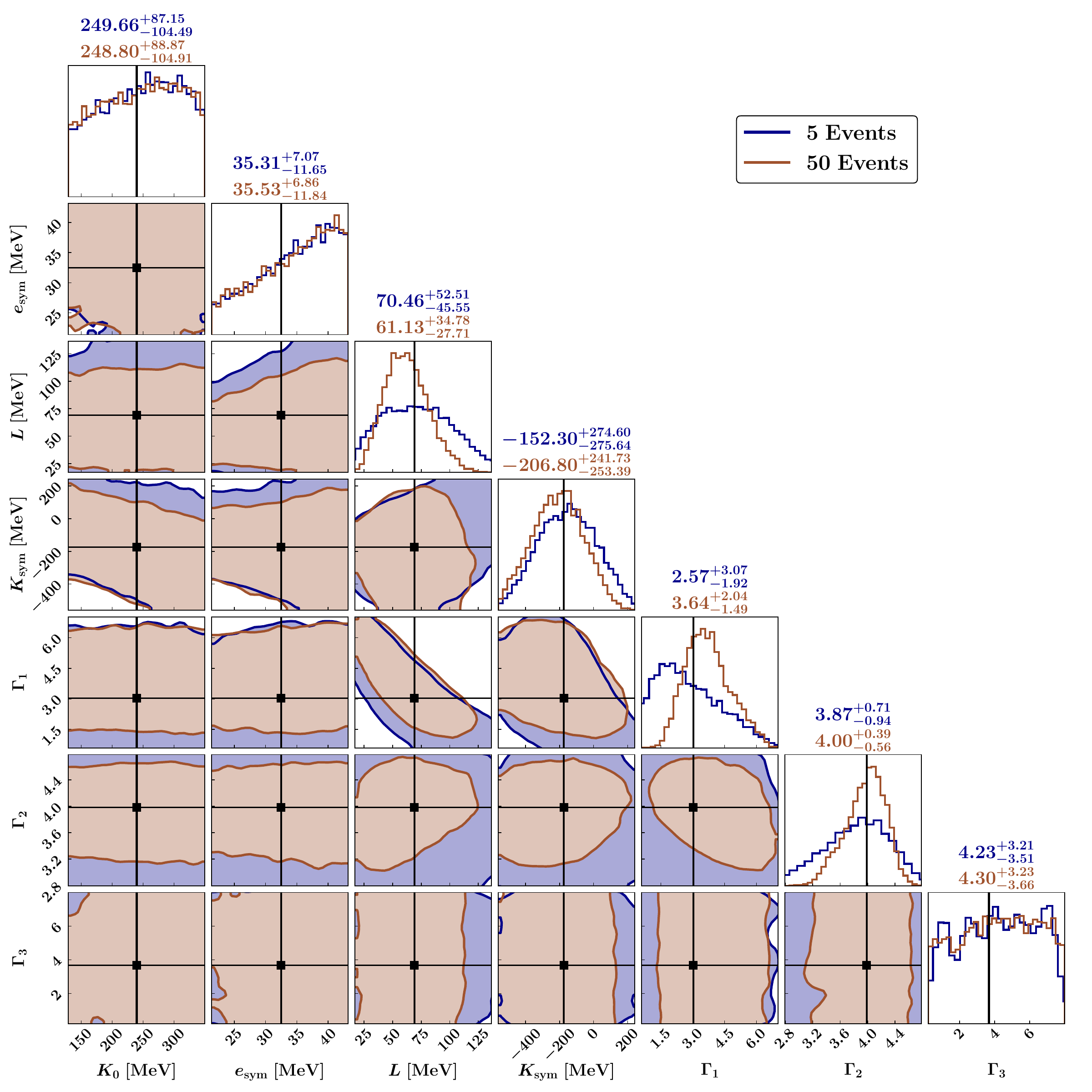}
    \caption{Same as Fig.~\ref{fig:EoS_corner_gaussian}, but BNSs follow the double Gaussian mass distribution.}
    \label{fig:EoS_corner_double_gaussian}
\end{figure*}

Comparison of NS EoS parameter constraints, inferred from $5$ and $50$ events, are shown in Figs.~\ref{fig:EoS_corner_gaussian}  and~\ref{fig:EoS_corner_double_gaussian}, corresponding to the Gaussian and double-Gaussian mass distributions, respectively.

\clearpage

\bibliography{reference}

\end{document}